\documentclass{aa}  
\usepackage{graphicx}
\usepackage{txfonts}
\usepackage{longtable}
\begin{document}
   \title{Pre-main sequence stars in open clusters}

   \subtitle{I.THE {\bf DAY-I} CATALOGUE }

   \author{Antonio J. Delgado\inst{1}, Emilio J. Alfaro
          \inst{1}
          \and
          Jo\~ao L. Yun\inst{2}
          }

   \offprints{A.J. Delgado}

   \institute{Instituto de Astrof\'\i sica de Andaluc\'\i a, CSIC, \\
    Apdo 3004, 18080-Granada, Spain\\
              \email{delgado@iaa.es},\email{emilio@iaa.es}
         \and
Centro de Astronomia e Astrof\'{\i}sica da Universidade de Lisboa,
Observat\'orio Astron\'omico de Lisboa, Tapada da Ajuda,
1349-018 Lisboa, Portugal \\
             \email{joao.yun@oal.ul.pt}
             }

   \date{Received September 15, 2006; accepted September 15, 2006}

 
  \abstract 
{We present the basic ideas and first results from the project we are 
  carrying out at present, the search for and characterisation of pre-main
  sequence (PMS) stars among the members of Galactic young
  clusters. The observations of 10 southern clusters, nine of them 
  located in the Carina-Sagittarius spiral arm of the Milky Way are 
  presented.} {We aim at listing candidate PMS member stars in young 
  clusters. The 
  catalogued stars will serve as a basis for future spectroscopic
  studies of individual objects to determine the properties of 
  stellar formation in the last phases before the main sequence
  stage. Properties such as the presence of residual envelopes or
  disks, age spread among PMS members, and the possible presence of
  several episodes of star formation in the clusters, are to be
  addressed.} {Multicolour photometry in the $UBVR_{C}I_{C}$ system has
  been obtained for 10 southern young clusters in the fourth Galactic
  quadrant, located between Galactic longitudes l=238$^o$ and
  l=310$^o$. For six clusters in the sample, the observations presented
  here provide the first published study based on CCD photometry. A
  quantitative comparison is performed with post-MS isochrones, and
  PMS isochrones from three different evolutionary models are used in
  the photometric membership analysis for possible PMS stars.}  {The
  observations produce photometric indices in the Johnson-Cousins
  photometric systems for a total of 26962 stars. The matching of
  our pixel coordinates with corresponding fields in the 2MASS data
  base provides astrometric calibration for all cataloged stars and
  $JHK$ 2MASS photometric indices for 60 \% of them. Post-MS cluster ages 
  range from 4 to 60 Myr, whereas the photometric membership 
  analysis assigns PMS membership assignment to a total of 842 stars, 
  covering an age range between 1 and 10 Myr. This information on the PMS
  candidate members has been collected into a catalogue, named DAY-I, which 
  contains 16 entries for 842 stars in the field of 10 southern clusters.} 
  {}


\authorrunning{Delgado et al.} 
\titlerunning{DAY-I Catalogue of PMS Stars}

   \maketitle
%

\section{Introduction}
Stars clusters are commonly accepted as the astrophysical objects best-suited 
to achieving reliable ages and distances of member stars at any location in 
the Milky Way. Furthermore, comparing photometric 
colour-magnitude (CM) diagrams with model  
isochrones simultaneously provides a way to estimate masses and chemical
compositions and to set constraints on the physics of the models themselves.
In particular, the systematic observation of the Galactic sample of young 
clusters offers a number of interesting possibilities for improving knowledge 
of how the star formation occurs on large scales in the Galaxy (e.g. Alfaro et 
al. 1991) and how stars form inside the clusters (e.g. Lada 2005, Lada \&
Alves 2004). By young clusters we understand objects with ages between 1 and 
10 Myr, which can be easily and productively observed at
both optical and infrared wavelengths and with telescopes around 1\,m in
diameter. For objects located not further than 3-4\,kpc, we are left with 
clusters containing observable PMS members, also 
called Post-TTauri stars (Mamajek et al. 2002), with spectral types 
from A to K, detectable in photometric diagrams deep down to $V$ 21-22, 
depending on reddening. These observations are feasible with small telescopes
and can be obtained for a wide sample of clusters 
in the nearest complexes of star formation, such as Carina and Cygnus, 
as well as for some clusters located at greater distances
in the direction of the Galactic anticentre.

These observations typically refer to older stars than those located in
embedded regions of 
star formation, such as the Orion nebula. But they still provide invaluable
information on the earliest phases of star formation. In particular, they
are expected to contain clues to the mass distribution of forming stars in 
selected regions, and they can 
also be studied with advantage at optical wavelengths. In this spectral 
region, the measurement of distance and age by means of ZAMS and post-MS 
isochrone fittings to the brightest part of the colour-magnitude diagram (CMD) 
is more reliable and leads to more accurate estimates of physical parameters 
for the member stars than for isolated objects.

PMS stars in clusters between 1 and 10 million yr are especially interesting
for investigating whether O-star formation terminates or, on the contrary, 
triggers the formation of less massive stars, as well as the possible 
deficit in the latter case(Sung et al. 1998). 
Furthermore, they trace the
evolution through the last phases of circumstellar disk dissipation and 
angular momentum variation of the central star (Stauffer et al. 1989; 
Haisch et al. 2001; Slesnick et al. 2002; Sicilia-Aguilar et al. 2005; 
Hartmann 2005). In particular, the proportion of CTTs to WTTs, possibly 
as a function of age, can be investigated (Herbig 1998). Recent work 
reveals the existence of long-lived disks around low-mass PMS stars, 
favourable for the formation of planetary disks (Lawson et al. 2004; 
Lyo et al. 2004), and a decrease in the time scale of disk dissipation 
with increasing mass (Hern\'andez et al. 2005). 

The presence of PMS members is addressed in most studies devoted to one or more
young clusters, but specific membership assignments are present only for close
or very bright objects. Ever since the first systematic investigation by 
Walker (1961, and references therein) of a group of bright young galactic 
clusters 
(NGC 2264, NGC 6530, IC5146, NGC6611), the presence of PMS stars among the
members of young clusters has been mentioned frequently in the literature. 
Many of these studies mention only the location of some stars on photometric 
diagrams, suggesting their PMS nature (FitzGerald et al. 1978; Forbes \& 
DuPuy 1978; Sagar \& Joshi 1979, 1981; Stone 1988; Forte \& Orsatti 1984; 
Witt et al. 1984; Heske \& Wendker 1984; Verschueren et al. 1990; 
Marschall et al. 1990; Munari et al. 1998; Carraro et al. 2001; 
Patat \& Carraro 2001).

A few studies include comparison with models and eventually more detailed 
observations of the most conspicuous PMS candidate members. Their main 
interest is focused on indications of coevality or non-coevality of the 
cluster members, the properties of the cluster mass function, and the possible 
peculiarities of the spatial distribution of the PMS member stars (Stauffer 
et al. 1989; Trullols \& Jordi 1997; Marco et al. 2001;  
DeGioia-Eastwood et al. 2001;  Marco \& Negueruela 2002). In some cases, 
membership probability is assigned on the basis of 
statistical considerations or discussed in some detail based on the 
global appearance of the CM diagrams as compared to PMS isochrones (Baade 
1983; Baume et al. 1999; Baume et al. 2003; Prisinzano et al. 2005)

On the other hand, detailed studies of the PMS candidate members, included in 
extensive analyses of the parent clusters, have been published mainly for a 
reduced collection of clusters (Th\'e et al. 1985; Hillenbrand et al. 1993; 
de Winter et al. 1997; van den Ancker et al. 1997; Sung et al. 1997; Herbig 
1998; Sung et al. 1998; Baume et al. 1999; Delgado et al. 1999; Sung et al.  
2000, van den Ancker et al. 2000; Park et al. 2000; Park \& Sung 2002; 
Delgado et al. 2004, 2006). Some of these studies use previously established 
PMS members, while others make their own PMS 
membership assignments, whereby the methods include location in several 
photometric diagrams, considerations of reddening, spectroscopic information 
for some stars, and values of specific photometric indices such as 
($R$-H$\alpha$).

Most recently, several published works contain  
exhaustive analyses of the PMS populations in regions close to the Sun, based 
on the widespread data sets available and including new observations (Mamajek 
et al. 2002; Luhman et al. 2003; Sartori et al. 2003; Pozzo et al. 
2003; Lyo et al. 2004; Prisinzano et al 2005). For these intensively studied 
objects, matches with available proper motions catalogues and lists of 
associated X-ray sources are used, and they provide the 
best-studied samples of PMS stars. Furthermore, in some special cases\,- 
clusters located at close distances, although not necessarily younger than  
10~Myr\,- detailed observations have been recently carried out to compare ages 
determined from isochrone fits, both for post-MS and for PMS models, with 
ages derived from the theoretical predictions on lithium depletion during the 
PMS evolution (Jeffries \& Oliveira 2005; Palla et al. 2005).

The conclusions about age spread, mass functions, and differences between
evolutionary models differ between studies. There is a general trend to admit
the existence of a significant age spread, which also produces age differences
as estimated from PMS and post-MS isochrones. Values suggested for age spread 
among PMS candidate members are roughly 5~Myr (Sung et al 2000; Park 
et al. 2000; Park \& Sung 2002), although values from 1 Myr (Hillenbrand et 
al. 1993) to more than 10~Myr (DeGioia-Eastwood et al. 2001; Pozzo et al. 
2003; Prisinzano et al. 2005) are claimed. On the other hand, results from 
different analyses of the closest associations, using a great wealth of 
available data offer different conclusions about the age differences between 
association members; see Mamajek et al. (2002) and Sartori et al. (2003) 
on the age spread in the Sco-Cen OB association. With respect to 
the mass function, some indications of a dependence of the results on the 
physical conditions of star formation is found in recent analyses (Luhman et 
al. 2003; Pozzo et al. 2003), whereas recent observations of PMS stars in the 
intermediate mass range in 30\,Dor (LMC) suggest no important differences 
between star formation in clusters in the Milky Way and outside it 
(Brandner et al. 2001).

In this paper we present the results for ten southern clusters, mostly located 
in the Carina-Sagittarius spiral arm of the Milky Way. This research is the 
continuation of a project, centred on the issues of interest already 
mentioned in 
this section, which has produced results for some clusters, mainly located in 
the Cygnus region and surroundings (see Delgado et al. 2004, and references 
therein). The main point in this research consists in assigning  
membership to individual candidate PMS stars on the basis of theoretical 
isochrone fitting on various CMD, in the same way as classical multicolour 
photometry is used to estimate membership of main sequence (MS) stars by 
means of ZAMS fitting. The procedure has been shown to perform better, when 
the reddening of the parent cluster is lower (Delgado \& Alfaro 2000). The 
project aims at obtaining homogeneous sample of PMS stars at different 
locations in the Galactic disk. It should provide a sample for future 
spectroscopic studies of individual objects, as well as for analysing (i) key 
issues of star formation in clusters, such as age and relative spatial 
distributions of MS and PMS members and (ii) the presence and dissipation 
time scales of circumstellar disks. Other points of interest are the 
assessment of an adequate T$_{eff}$-colour calibration for PMS stars and 
determining to what extent the application of calibrations valid for dwarfs 
is adequate for these stars (Delgado et al. 1998; Mamajek et al. 2002). Very 
interestingly,  Lyo et al. (2004) provide an observational isochrone from 
$\eta$-Cham observations, although for PMS stars with spectral types later 
than K7.

Section 2 contains a description of the observations and of the reduction and 
calibration procedures. A comparison with the previously published photometry 
of some of the clusters is presented. In Sect. 3 we describe and discuss the 
methodology used for determining the main physical properties of the host
clusters, as well as the membership assignments and ages for post-MS and
PMS stars in the observed area. Matches with the Two Micron All Sky 
Survey (2MASS)\footnote{http://www.ipac.caltech.edu/2mass/} catalogue 
and the resulting astrometry and near-infrared colour indices of the 
candidate members are addressed. Section 4 contains a brief summary of how 
the catalogue has been elaborated and a description of its contents.


\section{Observations}

The observations were carried out with the YALO\footnote{YALO is the 
{\bf Y}ale-{\bf A}URA-{\bf L}isbon-{\bf O}hio consortium (Bailyn et al. 1999). 
http://www.astronomy.ohio-state.edu/YALO/} telescope, operating at Cerro 
Tololo Observatory (CTIO), with the 
ANDICAM\footnote{http://www-astronomy.mps.ohio-state.edu/
$\sim$depoy/research/instrumentation/andicam/andicam.html} camera, in two
campaigns during December 2000 and May 2001. The optical channel of ANDICAM 
incorporates an array of 2024$\times$2024 pixels, with a plate scale of 0.3 
$\arcsec$/pixel. The $BVR_{C}I_{C}$ filters were mounted, together with the 
Str\"omgren ultraviolet $u$ filter, instead of the commonly-used $U$ passband 
from the Johnson system.

Several frames were secured through  each band for every cluster field 
(Table\,1), with integration times for short exposures of 5\,s for $BVRI$ 
bands and 10\,s for the $u$ band, and long exposures of 1200, 600, 300, 300, 
and 700\,s, for the $uBVRI$ bands, respectively. 
Three fields from the catalogue of standard stars by Landolt (1992) were also
observed. The fields of SA98-650, and Rubin~149, observed  in the 2000 
campaign, include 24 standard stars. The fields from the 2001 campaign, 
SA110-229, and PG1657+078, contain 8 standards. All these fields were observed 
every night at a minimum of three separated airmasses, in every band.
As mentioned above, the ultraviolet filter at our disposal was the $u$ 
Str\"omgren filter. No measurements of the standard fields were secured 
through this filter. The calibration of the $(U-B)$ colour index is feasible 
if we dispose of accurate published Johnson $(U-B)$ values in the field, and 
a reliable zero-point correction between instrumental and ``standard'' values 
can be determined. 

The different frames were reduced with the routines in the 
IRAF\footnote{The Image Reduction and Analysis Facility (IRAF) is distributed 
by the national Optical Astronomical Observatory, which is operated by the 
Association of Universities for Research in Astronomy, Inc. (AURA) under 
cooperative agreement with the National Science Foundation} package. Aperture 
corrections and atmospheric extinction corrections were applied to the 
magnitudes in each frame. Average extinction coefficients for 
CTIO\footnote{http://www.ctio.noao.edu/facil/s4/sec4.html\#b} were used.
Instrumental magnitudes were then calculated as weighted averages of the 
magnitudes in all frames. The weights used are the inverse of the squared 
PSF photometric errors.

The standard calibration of $UBV$ indices was carried out by direct correlation
of instrumental values with published photometry for the individual clusters, 
included in the WEBDA\footnote{http://www.univie.ac.at/webda//webda.html} 
database. These correlations produce accurate values for our purpose and 
offer a most reliable way of obtaining well-behaved $UBV$ indices (Delgado \& 
Alfaro 2000). The calibrations consist in determining zero-point corrections 
from the comparisons between instrumental and published values, with eventual 
colour-dependent terms after examining the residuals. The root-mean-squared 
deviation of the final residuals is adopted as the uncertainty on the 
calibration.
The calibration of the $(V-R)_{C}$ and $(V-I)_{C}$ indices was performed with 
the help of the observed Landolt standard stars, using the classical 
linear model for the calibration equations (see Delgado et al. 1998). As 
before, the root-mean-squared deviations of the fits were taken as the 
uncertainty on the calibrations.

The total errors assumed for the final standard indices are quadratic sums of 
the photometric errors and calibration uncertainties as described above. In 
Table 1, we list the target clusters' equatorial and galactic coordinates and 
the photometric study of reference used to compute the $UBV$ calibration. In 
Fig.\,1 we represent the variation in the uncertainties in a joint plot 
with $V$ for all the measured stars together. The different 
levels, appreciable at the brightest magnitudes, reproduce the different 
uncertainties of the calibration for different clusters.

\begin{table}
\caption{Observed clusters}
\label{table:1}
     $$ 
         \begin{array}{l r r r r c }
            \hline
            \noalign{\smallskip}
 $Cluster$ & \multicolumn{1}{c} {$RA$} & \multicolumn{1}{c} {$DEC$} & \multicolumn{1}{c} {$l$} & \multicolumn{1}{c} {$b$}  & $Ref$.^{\mathrm{1}}  \\
           & \multicolumn {2}{c} {$Epoch$~~2000}                    & \multicolumn {2}{c} {$deg$}  &             \\
            \noalign{\smallskip}
            \hline
            \noalign{\smallskip}
 $NGC 2367$      &  ~~7 ~20 ~04.54 & ~~-21 ~53 ~02.7 & ~~235.65 & -3.84  &  $a$  \\
 $NGC 3293$      & ~~10 ~35 ~48.77 & ~~-58 ~13 ~28.1 & ~~285.85 &  0.07  &  $b$  \\
 $Collinder 228$ & ~~10 ~43 ~01.26 & ~~-60 ~00 ~44.8 & ~~287.52 & -1.03  &  $c$  \\
 $Hogg10$        & ~~11 ~10 ~55.02 & ~~-60 ~22 ~18.1 & ~~290.80 &  0.10  &  $d$  \\
 $Hogg11$        & ~~11 ~11 ~32.63 & ~~-60 ~22 ~49.3 & ~~290.89 &  0.14  &  $e$  \\
 $Trumpler 18$   & ~~11 ~11 ~26.29 & ~~-60 ~40 ~19.1 & ~~290.99 & -0.14  &  $f$  \\
 $NGC 3590$      & ~~11 ~12 ~55.22 & ~~-60 ~46 ~30.7 & ~~291.21 & -0.18  &  $d$  \\
 $NGC 4103$      & ~~12 ~06 ~42.65 & ~~-61 ~14 ~26.8 & ~~297.57 &  1.18  &  $g$  \\
 $NGC 4463$      & ~~12 ~29 ~57.47 & ~~-64 ~47 ~55.7 & ~~300.65 & -2.01  &  $h$  \\
 $NGC 5606$      & ~~14 ~27 ~47.30 & ~~-59 ~38 ~34.8 & ~~314.84 &  0.99  &  $i$  \\
            \noalign{\smallskip}
            \hline
         \end{array}
     $$ 

\begin{list}{}{}
\item[$^{\mathrm{1}}$] 
{\bf a:} Vogt \& Moffat 1972. {\bf b:} Turner et al. 1980. {\bf c:} Carraro \& Patat 2001. {\bf d:} Moffat \& Vogt 1975. {\bf e:} Clari\'a 1976. {\bf f:} V\'azquez \& Feinstein 1990. {\bf g:} Wesselink 1969. {\bf h:} Moffat \& Vogt 1973. {\bf i:} V\'azquez \& Feinstein 1991.
\end{list}

\end{table}
   \begin{figure*}
   \centering
   \includegraphics[width=10cm]{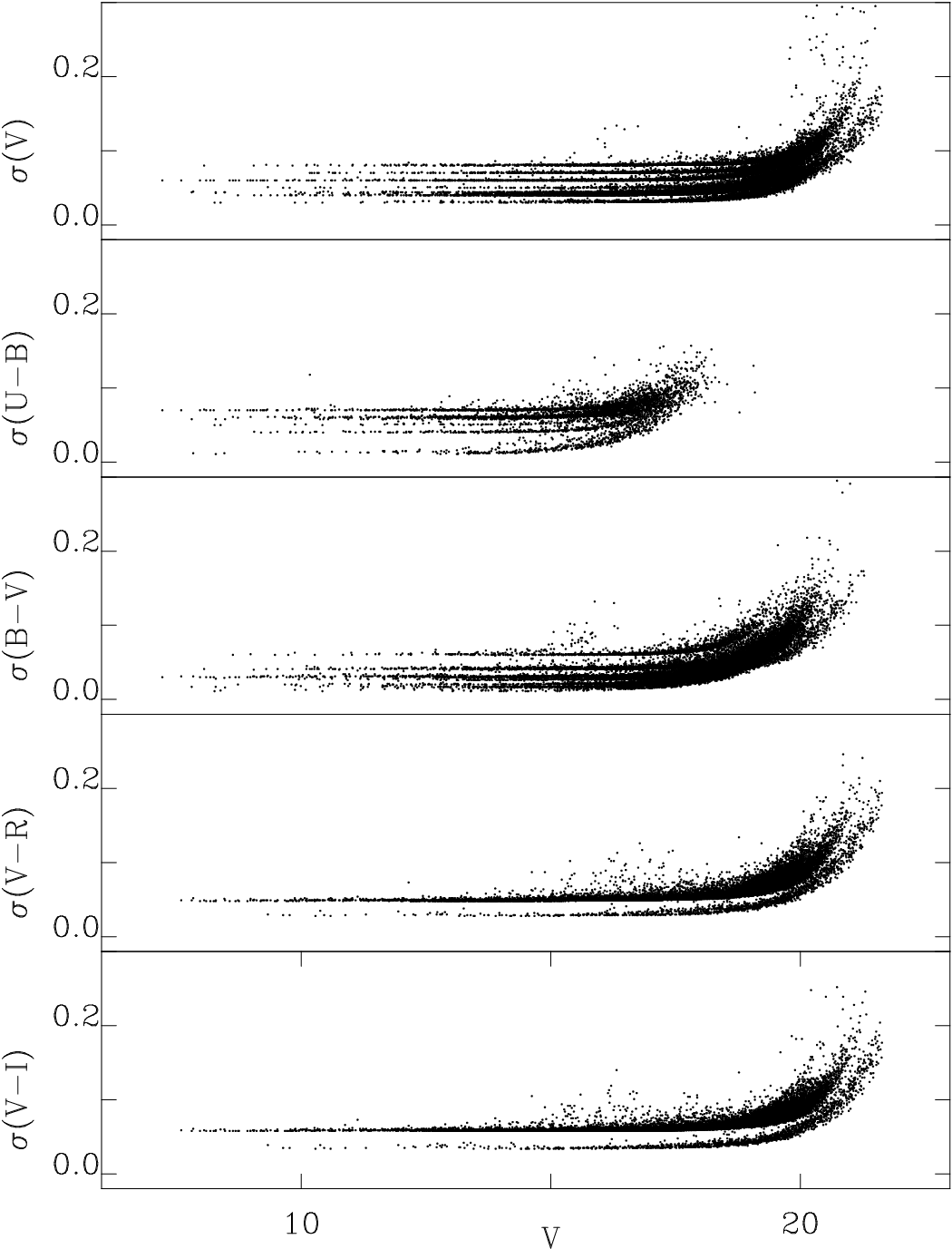}
   \caption{Variation with $V$ in the total uncertainties of standard indices
 for all observed clusters}
              \label{FigErr}%
    \end{figure*}
%

A comparison between our absolute photometry and those published for the 
clusters with CCD data, provides the values listed in Table 2. In this table 
we list the average differences in the first row, and the root-mean-squared 
(rms) deviations for each cluster in hundreths of magnitude in the second 
row. The high dispersion in some cases originates in the large errors for 
faint stars. We do, however, note the absence of significant systematic 
differences between our adopted colours and the published ones. Restricting 
the comparison to stars brighter than $V$=20, the rms deviation of the 
differences decreases below 0.1 in all cases.

\begin{table}
\caption{Comparison to published CCD photometry}
\label{table:2}
     $$ 
         \begin{array}{l r r r r r c}
            \hline
            \noalign{\smallskip}
 $Cluster$ & (V) & (U-B) & (B-V) & (V-R) & (V-I) & $Ref$.^{\mathrm{1}} \\

            \noalign{\smallskip}
            \hline
            \noalign{\smallskip}
$NGC 3293$      &  0.00 & -0.04 & -0.01 & -0.01 & -0.00  & $a$ \\
                &    19 &    11 &    11 &    21 &    10  &     \\
$NGC 4103$      &  0.04 & -0.03 & -0.02 &  0.04 & -0.03  & $b$ \\
                &    13 &    13 &    13 &    12 &    14  &     \\
$NGC 4103$      &  0.04 &       &  0.06 &       &  0.04  & $c$ \\
                &     9 &       &    12 &       &     8  &     \\
$NGC 5606$      &  0.02 &  0.01 & -0.02 &  0.06 &  0.05  & $d$ \\
                &    12 &    11 &    16 &    10 &     9  &     \\
$Collinder 228$ & -0.02 & -0.01 &  0.00 &  0.03 &  0.04  & $e$ \\
                &    11 &     8 &     9 &     8 &    11  &    \\
            \noalign{\smallskip}
            \hline
         \end{array}
     $$ 

\begin{list}{}{}
\item[$^{\mathrm{1}}$] 
{\bf a:} Baume et al. 2003. {\bf b:} Sagar \& Cannon 1997. {\bf c:} Sanner et 
al. 2001.{\bf d:} V\'azquez et al. 1994. {\bf e:} Carraro \& Patat 2001.
\end{list}

\end{table}

For Collinder\,228, the $UBV$ calibration in terms of published photoelectric 
values (Feinstein et al. 1976; Turner \& Moffat 1980) provides unreliable
results. Just for this cluster, the CCD values by Carraro \& Patat (2001) 
were then used to calibrate the $V, (U-B)$, and $(B-V)$ indices.

\section{Construction of the catalogue}

Our main objective is to detect ``bona fide'' PMS cluster-member 
candidates from multicolour photometry in the area of the cluster. Candidates
not only need to be assigned as cluster members, but also have to be located 
in a region of the CM diagrams, which are compatible with a PMS evolutionary 
stage. We therefore devote some room to explaining the methodology used to 
assign cluster membership to the individual stars and to distinguishing 
between 
post-MS and PMS members. This information can be obtained from the comparative
analysis of the observational CC and CM diagrams, and the comparison between 
observational and model diagrams. 

\subsection{Colour excess and distance: MS membership}

The first step in the procedure of membership assignement is the calculation 
of color excess and distance for MS stars. These values are to be used as 
reference to estimate the membership for PMS candidate-members. We note that 
most of the clusters in our sample do not have membership 
assignments for stars in the the region of the CMD where PMS members would be 
expected. Only NGC\,3292, the most-studied object, has membership assignments 
for stars fainter than $M_V$=2 (Baume et al. 2003). Some of our observed stars 
in Collinder 228 could be included in the $BVRIH\alpha$ photometric study 
of a stellar field in the Carina region by H\"agele et al. (2003). 
However, this study does not provide published or accessible data that could 
be compared to our results.

To calculate colour excess and absolute magnitude from the $(U-B),(B-V)$ 
CC diagram, we used the standard values of the reddening parameters, 
$\alpha\equiv E(U-B)/E(B-V)=0.72$, $R\equiv A_V/E(B-V)=3.1$. For NGC\,3293,
the detailed study of the reddening slope in the cluster field 
by Turner et al. (1980), suggests the value $\alpha=0.76$, so it has 
been adopted here.

\subsubsection{Selection of probable, unevolved MS members}

Our procedure for first selecting the upper MS probable members is to use the 
location of stars in the CM and CC diagrams, as described in Delgado et al. 
(1998). We start with the assumption that all clusters have a certain number 
of unevolved MS members with spectral types bluer than A0. A tentative 
selection of these stars is made visually in the $V,(B-V)$ CM diagram, also 
checking that the stars are congruently located in the other CM diagrams. This
procedure is shown in Fig.\,2, where the example of NGC\,2367 is plotted. For 
the stars tentatively selected (left panel) a colour excess and a 
corresponding absolute magnitude are calculated by means of the calibrated 
ZAMS line (Schmidt-Kaler 1982).

The quantity $V-3.1 \times E(B-V)$ is then plotted versus $M_V$, also 
considering previously published values of cluster parameters. In
the right panel of Fig.\,2, we see this plot for the tentative member stars, 
with the two different published values for the distance modulus of the 
cluster indicated by two different lines (Vogt \& Moffat 1972; Kharchenko 
2005, also adopted in WEBDA). In view of this plot, we select stars 
as probable unevolved MS members and others as evolved post-MS members, which 
will be used later to perform a quantitative estimate of the cluster age. The 
selection performed in these 
plots is based on how stars that evolved away from the ZAMS will show a 
lower distance modulus than the unevolved ones. In the same way, rotating or 
binary stars would have the same effect upon the DM calculation so are also 
rejected, as for the star marked with a triangle in the right panel of Fig.\,2.

   \begin{figure*}
   \centering
 \includegraphics[angle=-90,width=\textwidth]{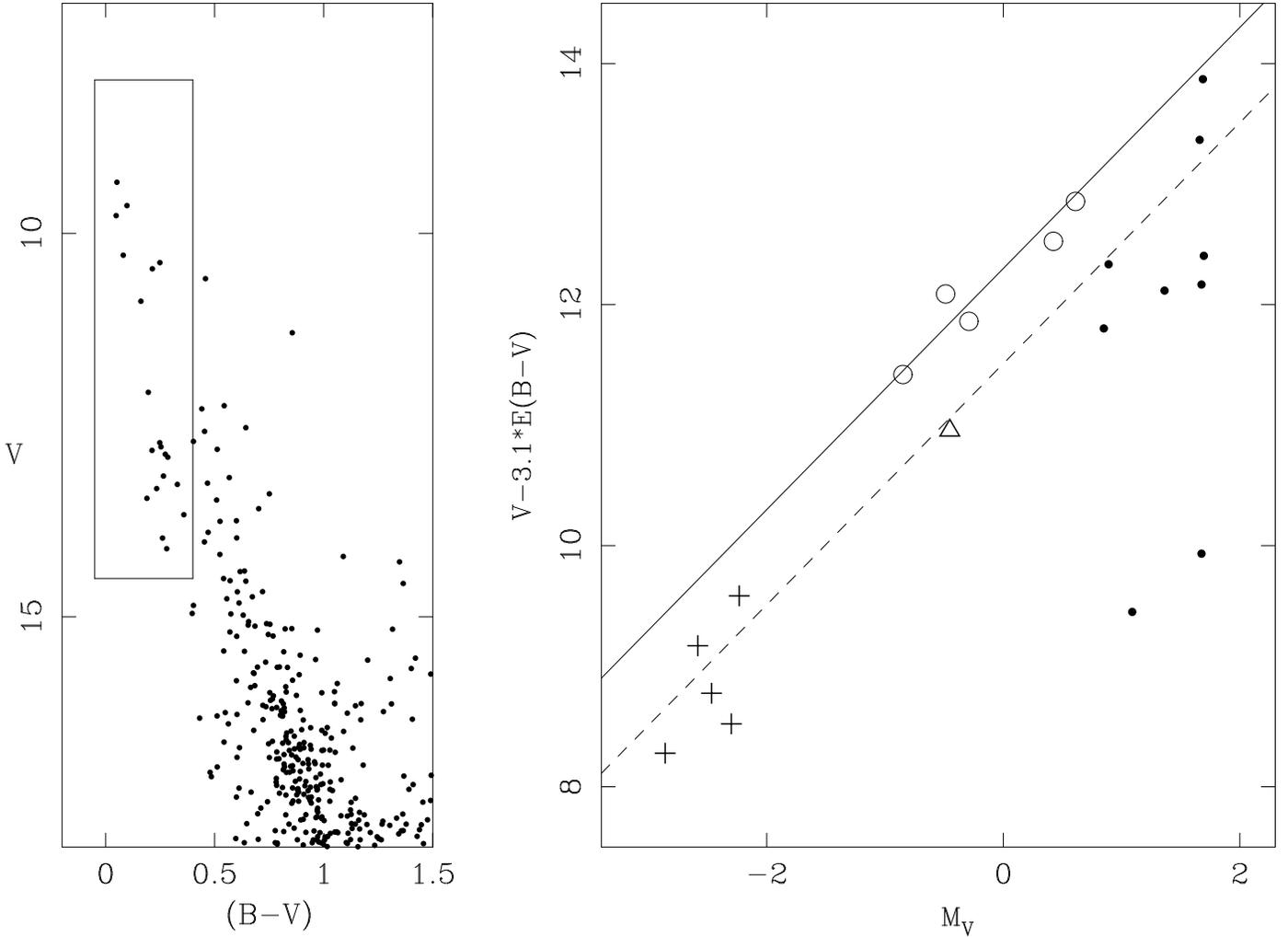}
   \caption{Example of selection of probable unevolved MS members. The left 
panel shows the tentative selection of upper MS members in the $V,(B-V)$ 
diagram of NGC\,2367 (see text). In the right panel, the quantity 
$V-3.1 \times E(B-V)$ is plotted for these stars versus $M_V$. The lines 
indicate published values of the distance modulus: Vogt \& Moffat 1972 
(continuous line) and Kharchenko 2005 (broken line). Circles are stars
selected from the plot as probable, unevolved MS members. Triangles and 
crosses are stars selected as post-MS members. Crosses mark the stars 
used to estimate the post-MS age of the cluster (see text)}
              \label{FigErr}%
    \end{figure*}

For most clusters in our sample, this procedure provides a reliable 
selection of upper non evolved MS members and post-MS members. For some of 
them, the selection of a cluster sequence is not unambiguous. In Fig.\,3 we 
show the case of Trumpler\,18 as an example, where two associations seem to 
overlap in the same field, at different distances. The selected possibility, 
according to previously published values, is shown with the larger distance 
modulus that would account for a star sequence located further away.

The selection of probable unevolved MS members is obviously of central 
importance for the membership analysis. The solutions adopted here, based on 
the plots like the one in Figs.\,2 and 3 are aimed at obtaining a basis for 
further PMS membership analysis, and they follow the appearance of our 
own CM diagrams, together with previously published results. On the other 
hand, there are interesting features in the CM diagrams of several clusters, 
suggesting the presence of several groupings at different distances in the 
same line of sight. This fact is apparent for lines of sight crossing the 
Carina spiral arm at different positions. These features can be guessed, for 
example, in the plots shown in both Figs.\,2 and 3. In this paper we only 
elaborate a catalogue of candidate PMS cluster members. 
The discussion of these features in the CM diagrams of the target clusters 
is postponed for a future paper.

   \begin{figure*}
   \centering
 \includegraphics[angle=-90,width=\textwidth]{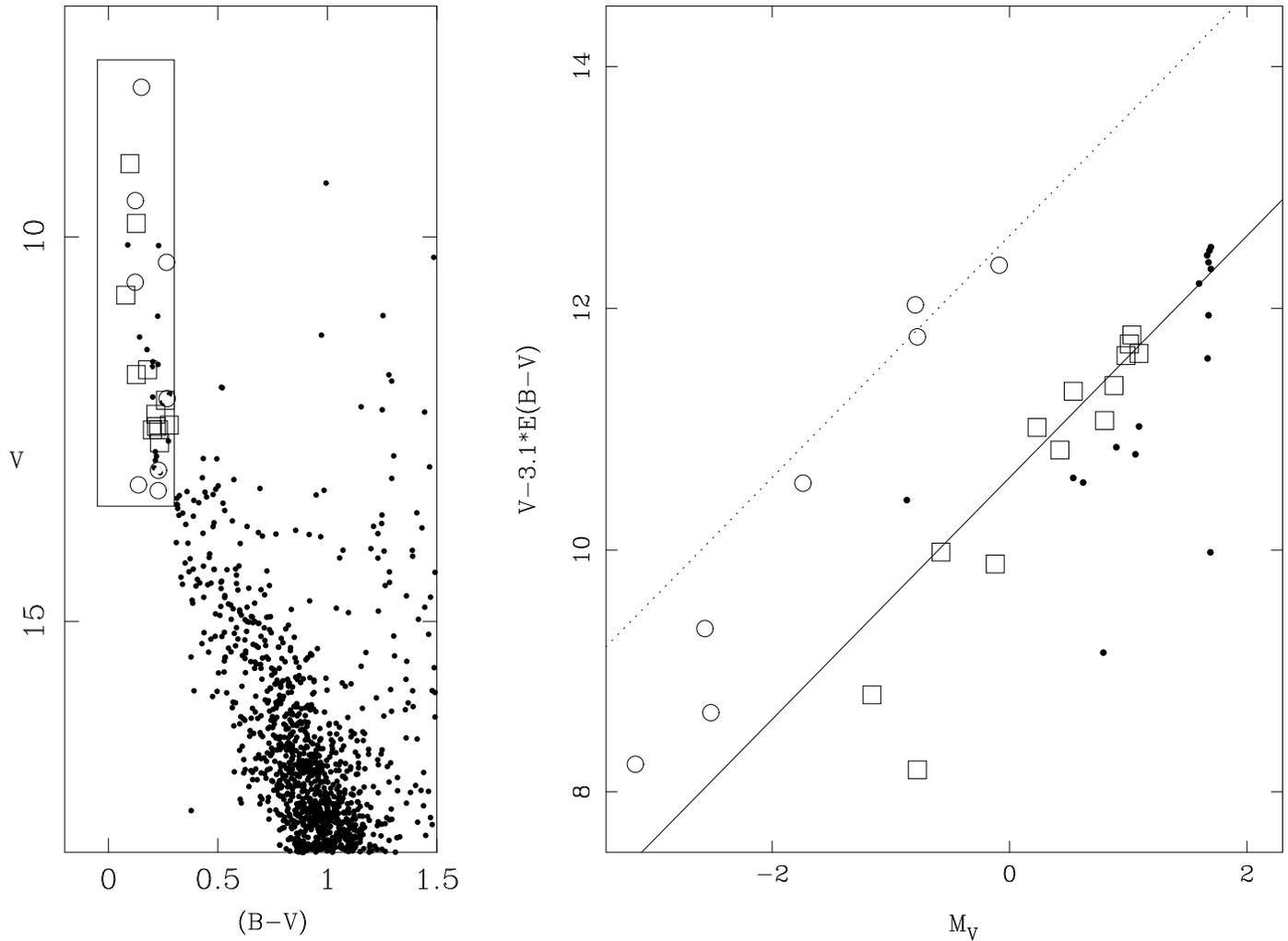}
   \caption{Same plots as in Fig.\,2, for Trumpler\,18. The stars in the 
two possible sequences are plotted with different symbols. The selection of 
upper probable MS cluster members is denoted with squares. The lines represent
distance moduli of 10.6 (continuous) and 12.7 (dotted)}
              \label{FigErr}%
    \end{figure*}

The method is applied to every cluster, and provides average values of colour 
excess and distance modulus for the probable unevolved MS members. These
are listed in Table\,3 with their rms deviations, together with the results 
derived by other authors. The values in Table\,3 agree in most cases with the 
published results. There are, however, some appreciable differences, as in 
the case of Hogg\,10, where the three published values of colour excess are 
noticeably higher than our value. This difference is due to our restricting 
the selection of probable unevolved MS members to 
avoid including binaries or evolved stars as much as possible. But 
we also generally obtain a more complete coverage of the cluster MS, 
whereas the photoelectric studies used contain only a few stars. In the 
particular case of Hogg\,10, for example, these are the ones with highest 
reddening.

\begin{table*}
\caption{Comparison with published results}
\label{table:1}
     $$ 
         \begin{array}{l r r r l}
            \hline
            \noalign{\smallskip}
  $Cluster$ & E(B-V) & DM  & Log(Age) &  Reference  \\
            \noalign{\smallskip}
            \hline
            \noalign{\smallskip}

 $NGC 2367$      & 0.33       &   11.51     &     6.74      &  $WEBDA$                   \\
                 & 0.35       &   12.28     &               &  $Vogt \& Moffat 1972$     \\
                 & 0.33       &   11.51     &     7.01      &  $Kharchenko et al. 2005$  \\
                 & {\bf 0.38 \pm 0.06} & {\bf 12.3 \pm 0.2} & {\bf 6.6 \pm 0.3} & $This work$ \\
 $NGC 3293$      & 0.26       &   11.83     &     7.01   &  $WEBDA$                      \\
                 & 0.29       &   11.99     &     6.70   &  $Turner et al. 1980$         \\
                 & 0.31       &   12.10     &     7.00   &  $Feinstein \& Marraco 1980$  \\
                 &            &   11.95     &            &  $Shobbrook 1983$             \\
                 & 0.25       & 11.8-12.6   &            &  $cited by  Shobbrook 1983$   \\
                 & 0.22       &   12.20     &     6.90   &  $Baume et al. 2003$          \\
                 & 0.25       &   11.96     &     6.94   &  $Kharchenko et al. 2005$     \\
                 & {\bf 0.29 \pm 0.04} & {\bf 12.0 \pm 0.2} & {\bf 6.8 \pm 0.4} & $This work$ \\
 $Collinder 228$ & 0.34       &   11.71     &     6.83   &  $WEBDA$                      \\
                 & 0.39       &   12.00     &     6.70   &  $Feinstein et al. 1976$      \\
                 & 0.32       &   12.17     &            &  $Turner \& Moffat 1980$      \\
                 & 0.30       &   11.39     &     6.90   &  $Carraro \& Patat 2001$      \\
                 & 0.37       &   12.45     &            &  $Massey et al. 2001$         \\
                 & 0.26       &   12.21     &     6.68   &  $Kharchenko et al. 2005$     \\
                 & {\bf 0.32 \pm 0.12} & {\bf 12.0 \pm 0.3} & {\bf 6.7 \pm 0.2} & $This work$ \\
 $Hogg 10$       & 0.46       &   11.25     &     6.78   &  $WEBDA$                      \\
                 & 0.46       &   11.71     &            &  $Moffat \& Vogt 1975$        \\
                 & 0.49       &   12.24     &     7.55   &  $Claria 1976$                \\
                 & {\bf 0.36 \pm 0.05} & {\bf 11.9 \pm 0.2} & {\bf 6.8 \pm 0.5} & $This work$ \\
 $Hogg 11$       & 0.32       &   11.78     &     7.08   &  $WEBDA$                      \\
                 & 0.32       &   11.81     &            &  $Moffat \& Vogt 1975$        \\
                 & 0.24       &             &     6.90   &  $Ahumada et al. 2001$        \\
                 & {\bf 0.30 \pm 0.05} & {\bf 11.9 \pm 0.2} & {\bf 6.7 \pm 0.3} & $This work$ \\
 $Trumpler 18$   & 0.32       &   10.66     &     7.19   &  $WEBDA$                      \\
                 & 0.29       &   10.54     &            &  $Moffat \& Vogt 1975$        \\
                 & 0.29-0.47  & 10.54-11.84 &            &  $Cited by Claria 1976$       \\
                 & 0.31       &   10.66     &     7.77   &  $Kharchenko et al. 2005$     \\
                 & {\bf 0.30 \pm 0.04} & {\bf 10.6 \pm 0.2} & {\bf 7.5 \pm 0.5} & $This work$ \\
 $NGC 3590$      & 0.45       &   11.09     &     7.23   &  $WEBDA$                      \\
                 & 0.51       &   11.57     &            &  $Moffat \& Vogt 1975$        \\
                 & 0.51       &   11.79     &     7.56   &  $Claria 1976$                \\
                 & 0.45       &   11.09     &     7.55   &  $Kharchenko et al. 2005$     \\
                 & {\bf 0.52 \pm 0.08} & {\bf 11.7 \pm 0.2} & {\bf 7.2 \pm 0.2} & $This work$ \\
 $NGC 4103$      & 0.29       &   11.06     &     7.39   &  $WEBDA$                      \\
                 & 0.31       &   11.51     &     7.48   &  $Sagar \& Cannon 1997$       \\
                 & 0.26       &   11.70     &     7.30   &  $Sanner et al. 2001$         \\
                 & 0.28       &   11.11     &     7.59   &  $Kharchenko et al. 2005$     \\
                 & {\bf 0.32 \pm 0.04} & {\bf 11.5 \pm 0.2} & {\bf 7.3 \pm 0.2} & $This work$ \\
 $NGC 4463$      & 0.43       &   10.11     &     7.51   &  $WEBDA$                      \\
                 & 0.44       &   11.77     &            &  $Moffat \& Vogt 1973$        \\
                 & 0.52       &   10.11     &     7.97   &  $Kharchenko et al. 2005$     \\
                 & {\bf 0.39 \pm 0.04} & {\bf 11.1 \pm 0.1} & {\bf 7.3 \pm 0.5} & $This work$ \\
 $NGC 5606$      & 0.47       &   11.28     &     7.08   &  $WEBDA$                      \\
                 & 0.49       &   11.40     &            &  $Moffat \& Vogt 1973$        \\
                 & 0.51       &   11.90     &     6.82   &  $Vazquez et al. 1994$        \\
                 & 0.31       &             &     6.65   &  $Ahumada et al. 2001$        \\
                 & 0.47       &   11.28     &     6.84   &  $Kharchenko et al. 2005$     \\
                 & {\bf 0.49 \pm 0.05} & {\bf 11.8 \pm 0.2} & {\bf 7.0 \pm 0.3} & $This work$ \\
            \noalign{\smallskip}
            \hline
         \end{array}
     $$ 

\end{table*}

\subsubsection{Membership assignment for possible MS members}

The average values of colour excess and distance modulus for the probable
unevolved MS members are used to estimate membership for the remaining stars.

Values of colour excess and distance modulus are calculated for each star by 
comparing their indices with the ZAMS reference line (Schmidt-Kaler 1982) in 
the $(U-B),(B-V)$ CC diagram. The representative point of each star is 
shifted in the CC diagram, with the adopted reddening slope, and surrounded 
by an ellipse of semiaxes given by the uncertainties in $(U-B)$ and $(B-V)$ 
(Fig.\,1). The minimum distance at every intersection of this ellipse with 
the reference line provides a coluor excess $E(B-V)$ and a particular value 
of the absolute visual magnitude $M_V$, wich are used to calculate a value for 
the intrinsic distance modulus. 

The calculation is visualised in Fig.\,4, using three stars in the cluster 
NGC\,3590 as an example. Figure\,4 shows how more than one solution is 
possible for each star, depending on its location in the CC diagram. A star 
is then considered as a member if any one of the pairs of values $E(B-V),DM$ 
coincides, 
within the errors, with the average values from the probable unevolved MS 
members. In this comparison, the errors used are squared quadratic sums of the 
uncertainties plotted in Fig.\,1, and the rms deviation of the average 
values (Table\,3). In practice, when selecting MS members, we require that the 
colour excess be larger than the average value of the probable unevolved MS 
members minus the error and smaller than this value plus twice the error. 
When selecting PMS members, we simply require the color excess be above the 
value of the probable unevolved MS members minus the error. We try in this way 
to favour the detection of possible members, PMS members in particular, which 
are more affected by reddening than the probable unevolved MS members. 

We note that a star may already have one calculated value for the colour 
excess 
that fulfils the requirements for membership, prior to the distance modulus 
comparison. In considerations below we refer to these stars as reddening 
candidates. In fact, this constitutes the first membership filter. Obviously, 
all stars assigned as members are first reddening candidates, but only stars 
that fulfil this condition and the agreement of the corresponding distance 
modulus are kept as candidate members.

   \begin{figure*}
   \centering
 \includegraphics[angle=-90,width=\textwidth]{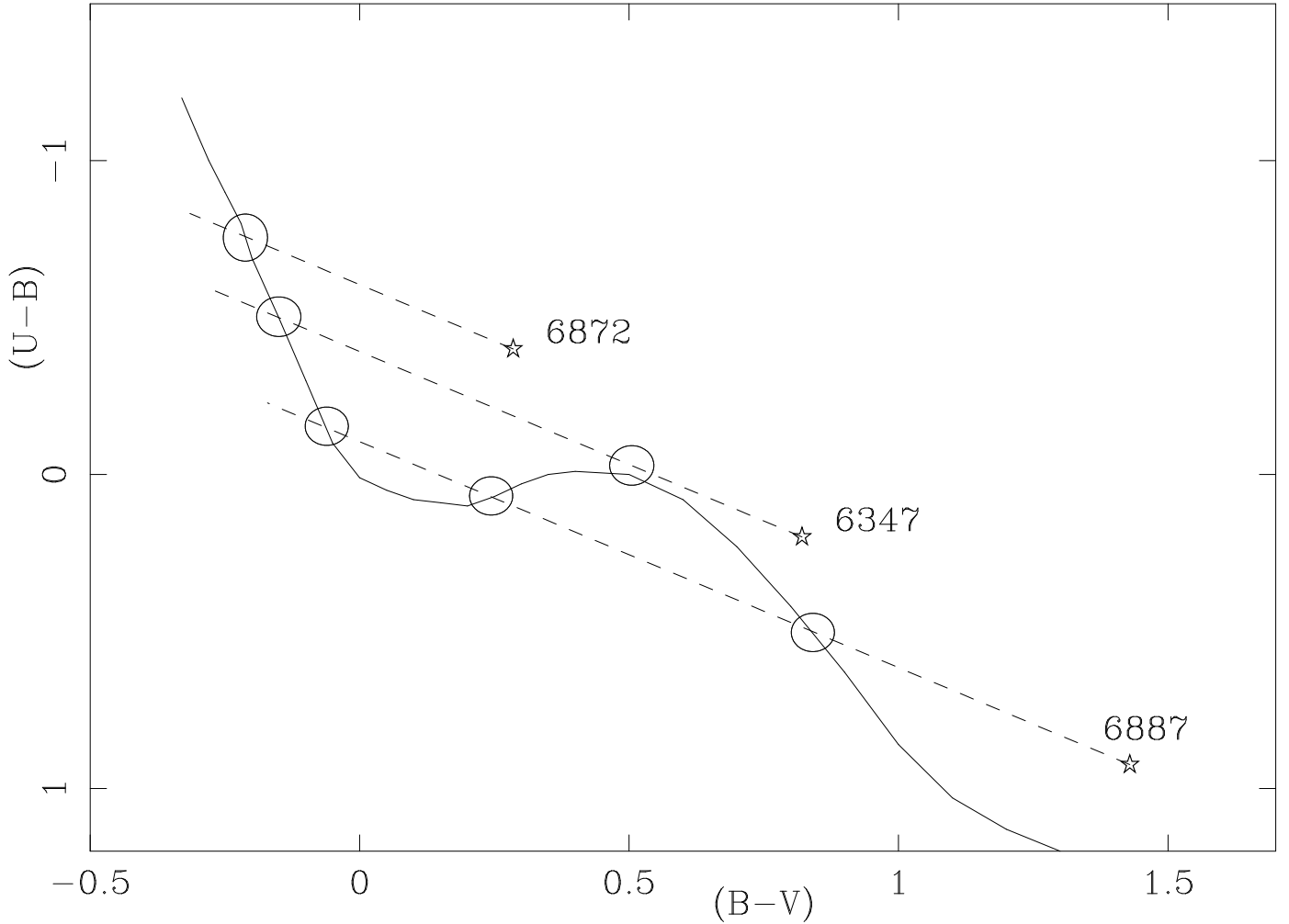}
   \caption{Example of the colour excess calculation in the $(U-B),(B-V)$ 
diagram. Three stars in NGC\,3590 with different types of solutions are 
plotted, together with the ZAMS (continuous line). Broken lines represent the 
shifts with the slope $E(U-B)/E(B-V)=0.72$. Ellipses represent the photometric 
uncertainties of the respective points, plotted in Figure\,1. }
              \label{FigErr}%
    \end{figure*}
%

\subsubsection{Age estimate for post-MS stars}

The age of the cluster is estimated from comparing of the  CM diagrams 
with theoretical post-MS isochrones. A simple quantitative procedure is used,
which provides a systematic age estimate instead of the commonly-used guess
from a visual comparison between data and isochrones. A similar procedure has 
been used by Trullols \& Jordi (1997). 

As done for the probable unevolved MS members, the post-MS probable 
members used in the calculations are selected from plots like those shown in 
Figs.\,1 and 2, also considering the consistent location of every star in 
all CM diagrams, and helped by results from previous studies. As is known, 
in addition to observational errors and to metallicity variations, which are 
not expected in the studied young clusters, the presence of rotation and 
binarity cause a broadening of the cluster's main sequence (see for instance 
Fernandes et al. 1996). This would lead to overestimating the cluster age. In 
our selection we try to avoid stars that could be affected by these effects. 

In the calculation we use isochrones from the Padova group (Girardi et al. 
2002), for LogAge(yr) values of 6.6, 6.7, 6.8, 6.9, 7.0, 7.2, 7.4, and 8. The 
isochrones are first shifted in all CM diagrams to account for the colour 
excess and distance moduli calculated for the probable unevolved MS members. 
Furthermore, the colour calibration used to translate the theoretical tracks
to the observational plane present offsets with respect to the ZAMS of 
Schmidt-Kaler (1982) or Kenyon \& Hartmann (1995), which coincide quite 
closely. When using the isochrones for age estimates, these 
offsets are previously corrected. 

For each involved star, the difference in colour with respect to every 
isochrone is measured in the four CM diagrams available. These differences 
are then either interpolated or extrapolated to the age value that would  fit 
the star position exactly. This provides a nominal age value for each 
star, and all these values are then averaged to give the post-MS age estimate 
for the cluster. The average values and their rms deviations are listed in 
Table\,3, together with other clusters parameters, and compared to published 
values. 

\subsection{PMS membership}

We distinguish between stars for which a colour excess can be calculated
in the CC diagram and stars that do not have this possibility, either because
of lacking $U$ measurement, the most common case, or because the shifting in 
the CC diagram does not provide any valid intersection with the reference 
lines. 

For stars with $(U-B)$ measurement, the calculation proceeds as explained 
above. These stars were also compared with the PMS isochrones, taken from 
three different models (D'Antona \& Mazzitelli 1997; Palla 1999; Siess et al. 
2000, respectively referred to in the following as D97, P99, and S00). The 
$(U-B)$ vs $(B-V)$ relation for these isochrones was the same as the one used 
for the ZAMS (Kenyon \& Hartmann 1995). The colour excesses were therefore the 
same, whereas the distance moduli were different, depending on the model and 
on the age of the isochrone. We used model isochrones of ages from 1 to 10\,Myr
in steps of 1\,Myr for the S00 models and 2\,Myr for D97 and P99 models. 
Therefore, we obtained up to 20 possible DM values, from 10 S00 isochrones, 5 
P99 isochrones, and 5 D97 isochrones. A star was then assigned as a PMS member 
if, in addition to the colour excess requirement for PMS candidates (Sect. 
3.1.2), one or more of these DM values agreed within the errors with the 
average value of the probable MS unevolved members. Obviously, this 
membership assignment wss done simultaneously in the two CM diagrams $V,(U-B)$ 
and $V,(B-V)$.

For stars without a valid $(U-B)$ value, the $CC$ diagram cannot be used, 
and we do not have calculated values of colour excess and distance modulus to 
compare with reference lines. For these stars, only PMS membership was
investigated. To assign PMS membership to these stars, we assumed every star 
was affected by the colour excess calculated for the probable unevolved MS 
members. To compare stars with isochrones in the $V,(V-R)$ and $V,(V-I)$ CM 
diagrams, we used reddening slopes for adequate achieving of a good ZAMS 
fitting of the probable unevolved MS members in the respective diagrams. 
These values for 
the reddening slopes are in good agreement with the values from a standard
reddening law (Cardelli et al. 1999). The PMS isochrones are shifted with these
values in the CM diagrams, and the comparison with the data gives a distance 
modulus for every star relative to each isochrone. A star has a membership 
assignment (with respect to an isochrone) if the obtained distance modulus 
coincides, within the errors, with the average value for the probable MS 
unevolved members. In this procedure, a star can have a membership assignement 
in one, two, or all three CM diagrams involved.

To clarify this procedure of member selection, we show examples in the CM 
diagrams of NGC3293 in Fig.\,5  of stars with different qualities of 
membership
assignment. The ellipses around each point reproduce the errors explained 
above, used for comparison: squared quadratic sums of the uncertainties 
plotted in Fig.\,1, and rms deviation of the average colour excess and 
distance modulus for the probable unevolved MS members. The isochrones 
plotted in this example are from  Siess et al. (2000).

   \begin{figure*}
   \centering
 \includegraphics[angle=-90,width=\textwidth]{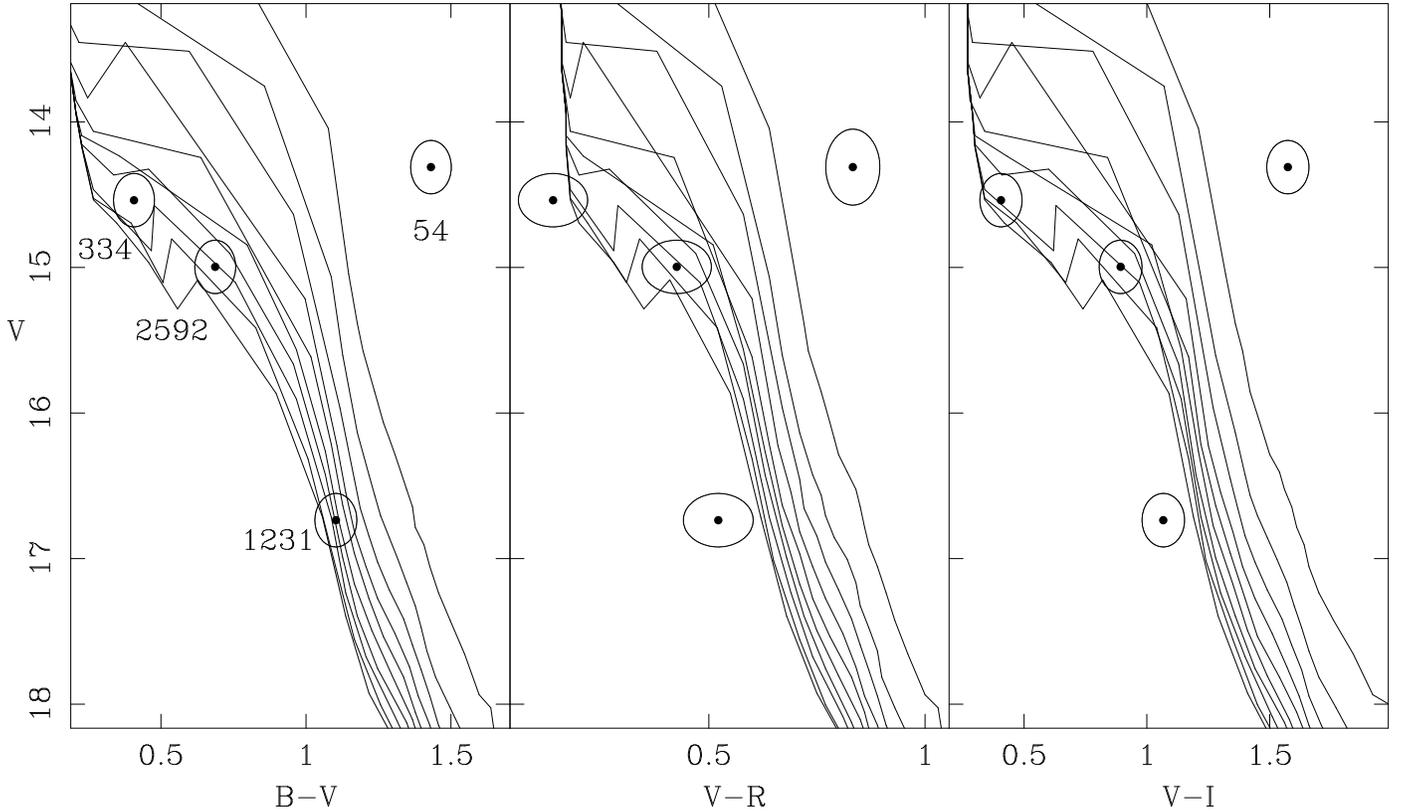}
   \caption{Example of PMS members selection in the CM diagrams of NGC\,3293. 
Star \# 54, non member; \#1231, member in the $V,(B-V)$ diagram; \#334, 
member in three CM diagrams, with different ages; \#2592, member in all three 
diagrams, with age differences among them below 1\,Myr. The axes of the 
ellipses around each point reproduce the squared quadratic sums of final 
uncertainties (Fig.\,1) and the rms deviation of cluster parameters for the 
probable unevolved MS members (Table\,3). Isochrones of ages 1 to 10 Myr from 
Siess et al. (2000) are plotted}
              \label{FigErr}%
    \end{figure*}

In both methods of membership assignement, the errors used in the comparisons 
are the squared quadratic sums of the uncertainties plotted in Fig.\,1, and 
the rms deviation of the average values for the probable unevolved MS members.
As can be appreciated in the example of Fig.\,5, the errors are usually 
larger than the separation between two consecutive isochrones. As a result, 
every star can be a candidate member with respect to several PMS isochrones. 
The average value of the corresponding ages is taken as the age of the PMS 
candidate. The average value of the ages for all assigned members and its 
rms value are adopted as the age and uncertainty of the PMS sequence of the 
cluster, as obtained from the particular CM diagram and model being compared.

Two more points have to be covered before finally selecting the candidate
PMS members. First, the procedure described above for analysing the membership
of stars without a valid $(U-B)$ index was also applied to stars with a 
valid $(U-B)$ value, {\sl and} qualified as reddening candidates (see the last
paragraph in Sect.\,3.1.2). For stars without a $(U-B)$ index, the average 
colour excess of the probable unevolved MS members was assumed, as explained 
above. For reddening candidates, their own calculated color excess value was 
used instead. In this way, PMS membership assignment can be established for a 
star with respect to several combinations of three CM diagrams, and even with 
respect to four CM diagrams. This is indeed the case of a star with an 
eventual assignment based on its calculated colour excess and distance 
modulus (as mentioned above, this obviously means simultaneous assignment in 
the $V,(U-B)$ and $V,(B-V)$ CM diagrams), and also assigned as a member from 
its location in the $V,(V-R)$ and $V,(V-I)$ CM diagrams. 

Second, the procedure for membership assignment may lead, for some stars with
a valid $(U-B)$ value to an assignment both as MS and PMS members. This will 
happen for stars located in the region of the CM diagrams where the ZAMS and 
the oldest PMS isochrones merge. In these cases, we adopted the PMS 
assignment for stars with spectral types later than A0 ($M_V$=1.3 on the 
ZAMS) and the MS assignment for stars earlier than this.
 
The final selection of candidate PMS members included only stars whith a 
membership assignment in at least three CM diagrams. Furthermore, to achieve 
a conservative estimate, we considered a further level of membership quality, 
with two more restrictions. A star is definitely considered as a possible PMS 
member if, in addition to being assigned in three CM diagrams, (i) its 
photometric
errors (PSF errors) are below 0.05 in all five $UBVRI$ bands and (ii) the ages 
obtained from all three (or eventually four) CM diagrams coincide within 
a range of 1\,Myr. The resulting assigned members are used to calculate the 
PMS age, as explained, and are the stars listed in our final catalogue.

The results of the selection procedure are shown in the CM diagrams of the
clusters. Figures\,6 to 10 show the $V,(B-V)$ CM diagrams for all clusters, 
compared to both post-MS (G02) and PMS (S00) isochrones. As mentioned before, 
three PMS model isochrones 
are used to calculate this membership assignment (D97, P99, and S00). Three 
different assignments and ages are therefore obtained for the PMS candidates 
in each cluster. In Figures\,6 to 10, we only plot the candidate PMS members 
with respect to S00 isochrones. In our final catalogue, we include assigned 
members with respect to any of the three model isochrones. In this respect, 
we have to mention that D97 isochrones cover only the faintest part of the CM 
diagrams, and they do not provide PMS membership assignment for stars earlier 
than G0.

   \begin{figure*}
   \centering
 \includegraphics[angle=-90,width=\textwidth]{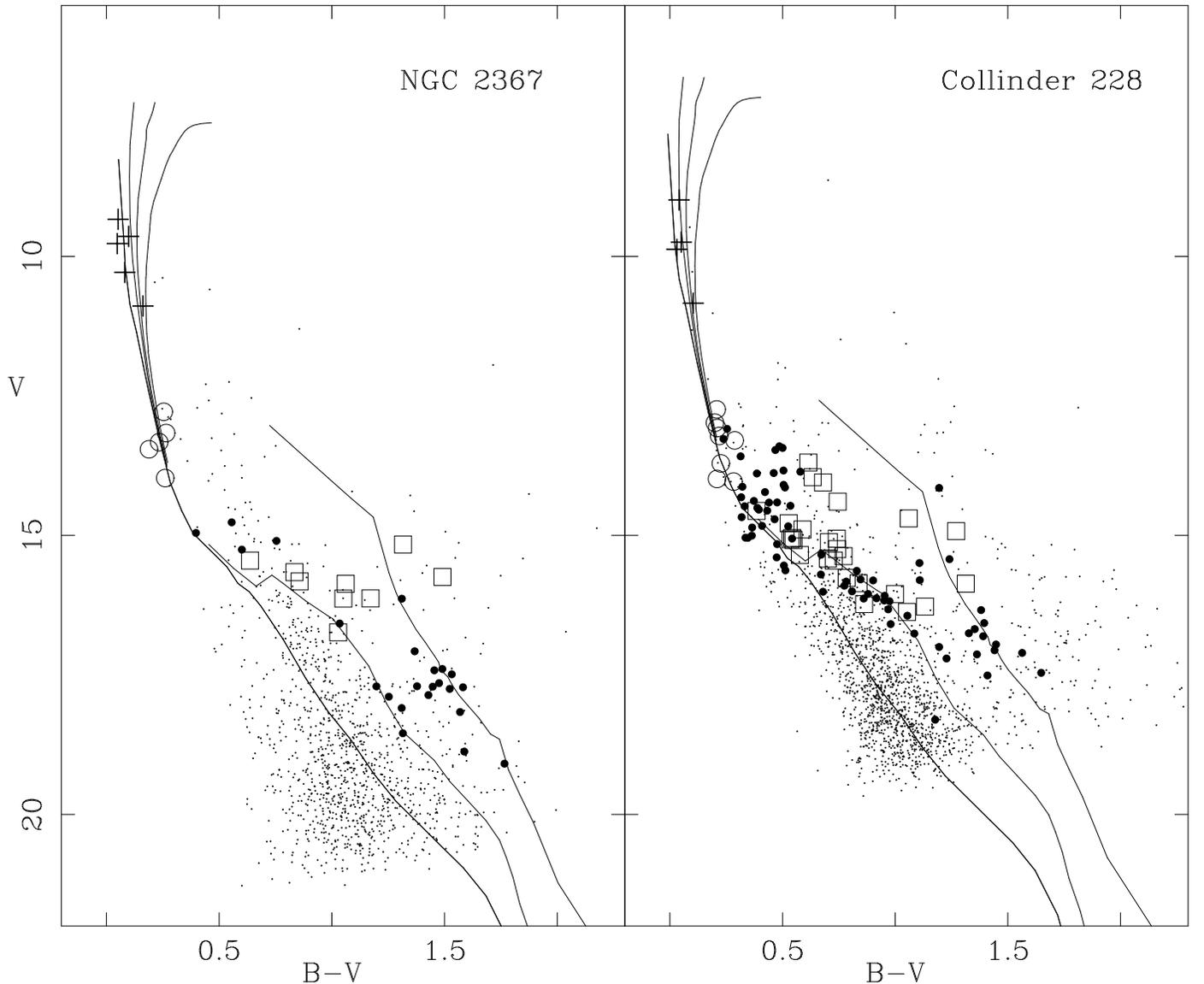}
   \caption{$V,(B-V)$ CM diagrams of NGC\,2367 and Collinder\,228, two of the 
youngest clusters in our sample. Circles represent the probable unevolved MS 
members. Crosses are the post:MS probable membrs selected to for the 
quantitative age estimate. Large dots represent PMS members in respect to any 
of the three model isochrones used, and selected from three CM diagrams.
Squares are PMS members assigned in all four CM diagrams. Small dots are the
remaining observed stars. The ZAMS line, post-MS isochrones by G02, for 
LogAge(yr) 6.6, 7.0, and 7.4, and S00 PMS isochrones for LogAge(yr) 6 and 7, 
are plotted for reference.}
              \label{FigErr}%
    \end{figure*}
%

   \begin{figure*}
   \centering
 \includegraphics[angle=-90,width=\textwidth]{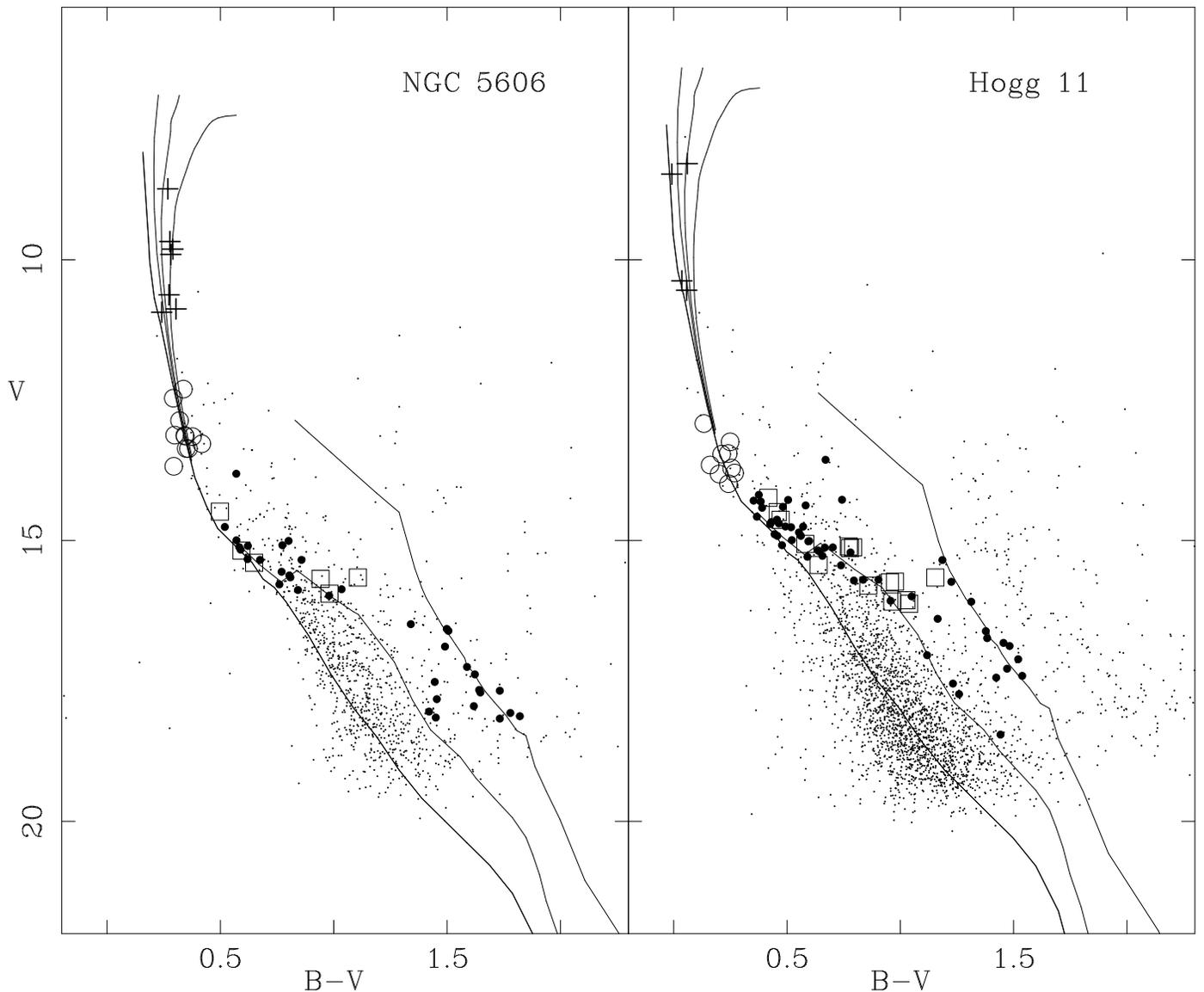}
   \caption{The same as Figure\,3, for the clusters NGC\,5606 and Hogg\,11.}

              \label{FigErr}%
    \end{figure*}
%

   \begin{figure*}
   \centering
 \includegraphics[angle=-90,width=\textwidth]{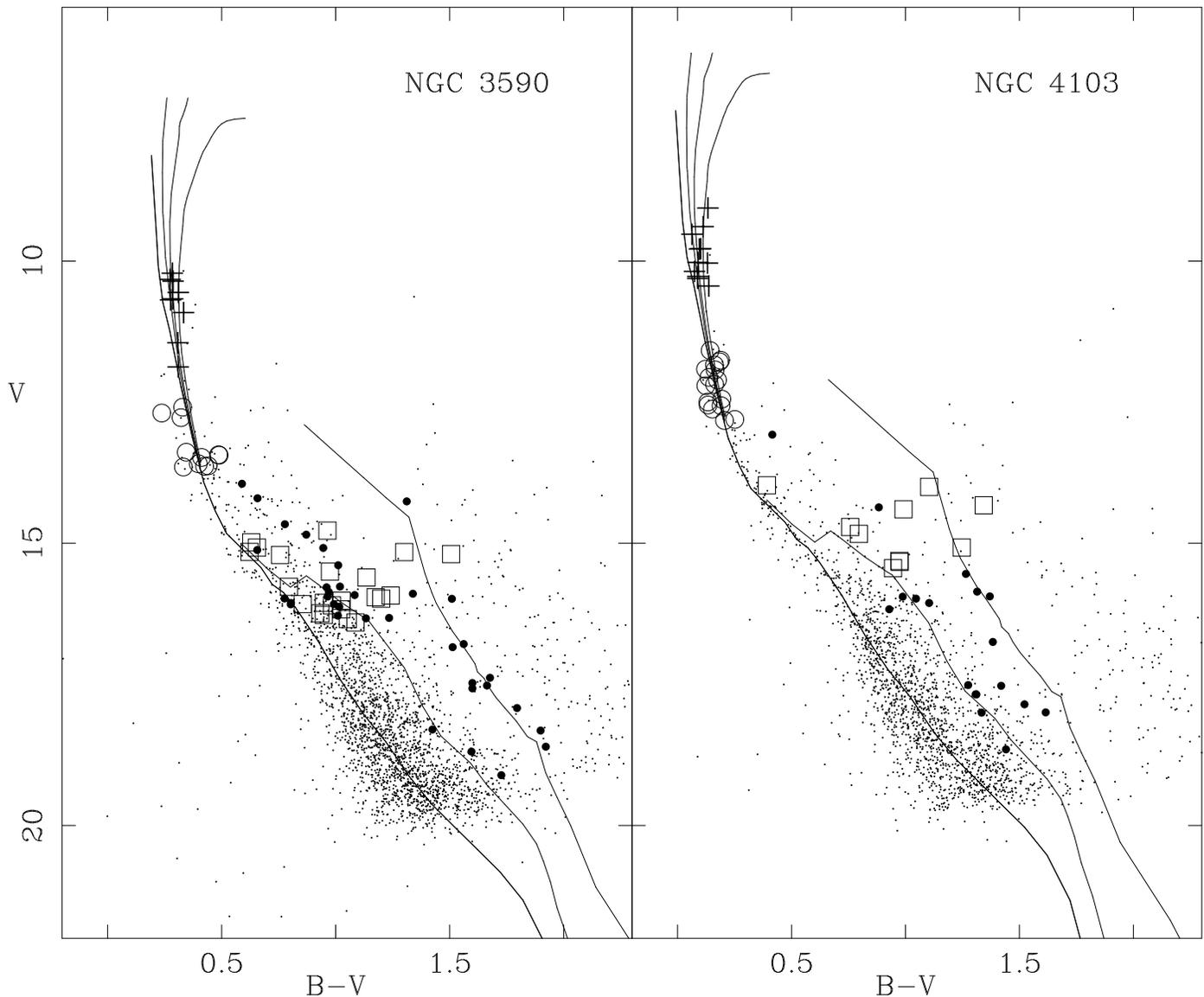}
   \caption{The same as Figure\,3, for the clusters NGC\,3590 and NGC\,4103.}

              \label{FigErr}%
    \end{figure*}
%

   \begin{figure*}
   \centering
 \includegraphics[angle=-90,width=\textwidth]{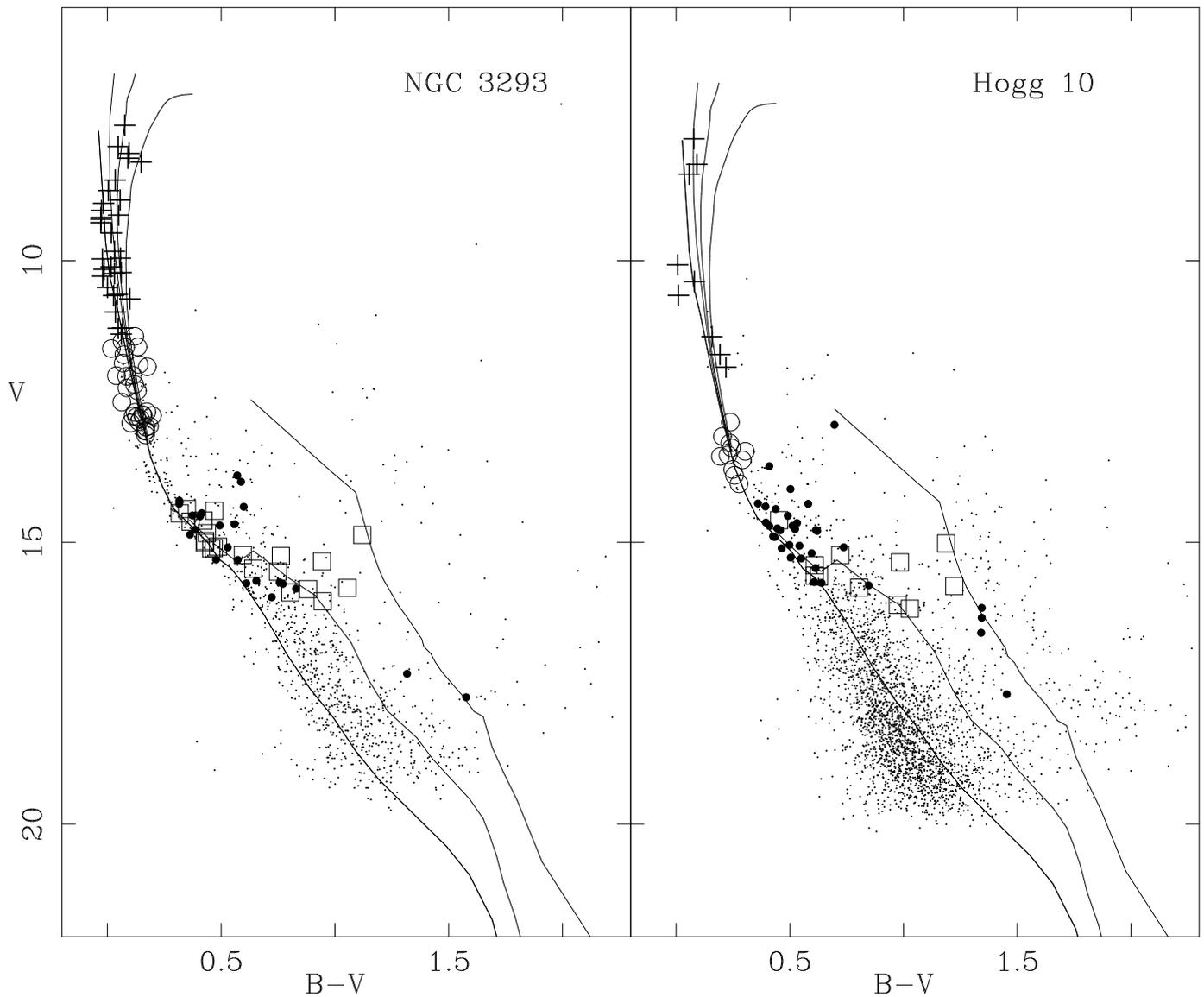}
   \caption{The same as Figure\,3, for the clusters NGC\,3293 and Hogg\,10.}

              \label{FigErr}%
    \end{figure*}
%

   \begin{figure*}
   \centering
 \includegraphics[angle=-90,width=\textwidth]{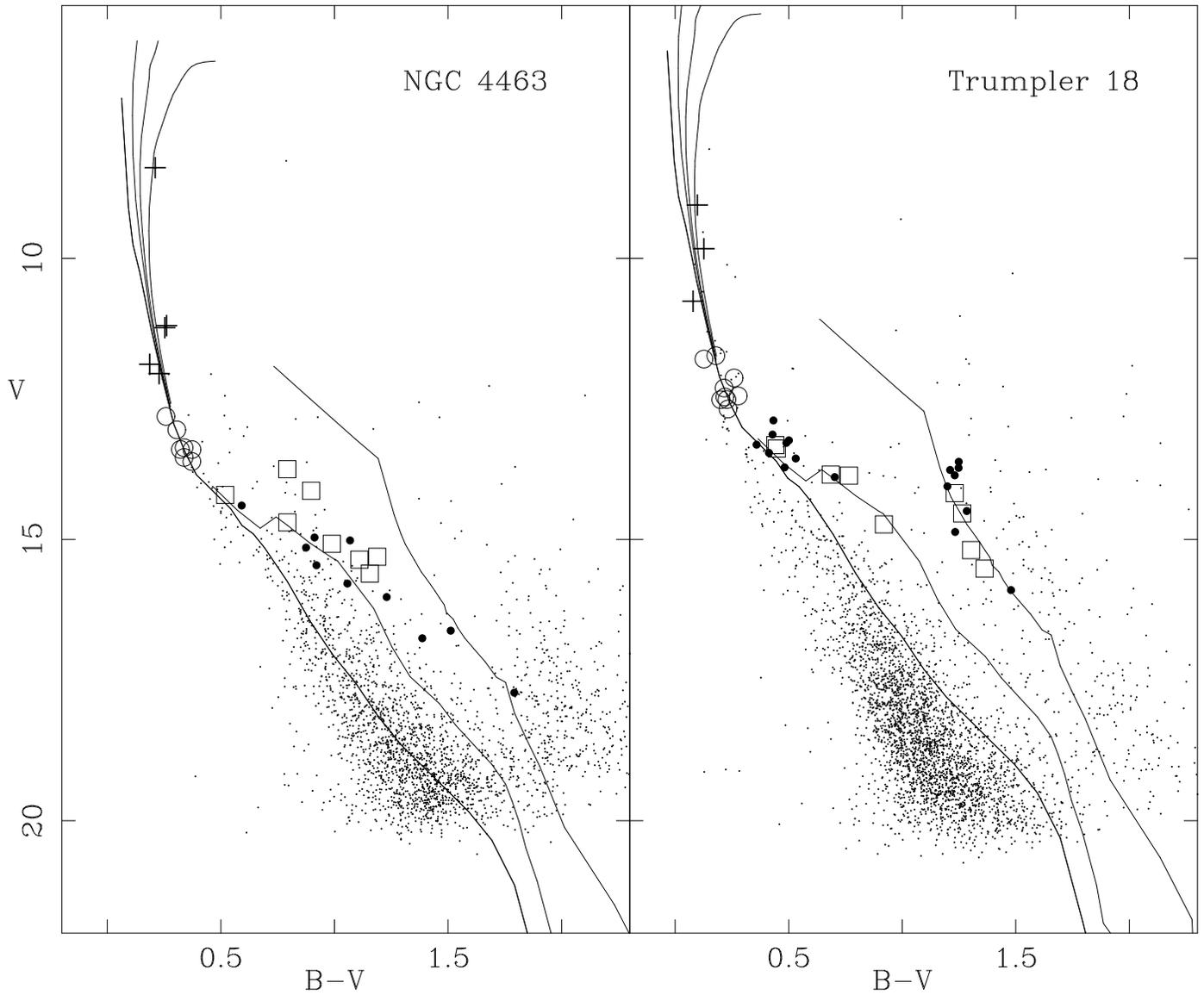}
\caption{The same as Figure\,3, for the clusters NGC\,4463 and Trumpler\,18.}

              \label{FigErr}%
    \end{figure*}
%

\subsection{Astrometry and 2MASS infrared data}

The pixel coordinates of our catalogue were transformed to equatorial 
coordinates with the IRAF tasks {\tt ccmap} and {\tt xy2rd}, using matching 
files with stars in common between our fields and corresponding fields in 
the 2MASS data base. The matching provides equatorial right ascension and 
declination for all the stars observed, and $JHK$ photometry for 15850 stars,
while 68\% of them have photometric quality flag A to D in all three $JHK$ 
bands (see the catalogue description 
at {\tt http://www.ipac.caltech.edu/\-2mass\-/releases/\-allsky/\-doc/}). 

The values from the 2MASS catalogue were transformed to the system of Bessel 
\& Brett (1988) with relations given by Carpenter (2001). These values are 
also included in the final catalogue. 

\subsection{Additional remarks}

Cluster membership assignments derived from kinematic studies have always been 
considered more reliable than those obtained from photometric analysis.
However, the determination of probable cluster members from analysing  
their location in the photometric diagrams has an ample and successful 
history, ever since its origin around the middle of the past century. We have 
profited from this experience, and with addition of new PMS evolutionary 
models, developed a methodology based on analysis of colour-colour and
colour-magnitude diagrams to obtain a reliable sample of PMS cluster candidate
members. This method was designed and developed in previous works and applied 
to several clusters (Delgado et al. 1998, 2000, 2004, 2006). Here, we use it 
to study a cluster sample, with homogeneous $UBVR_{C}I_C$ photometry, 
obtained at the same telescope and with the same instrumentation for all of 
them. The clusters are located mainly in the Carina complex of star formation 
(Alfaro et al.\, 1992), and they could therefore contain valuable information 
on the primary evolutionary stage of stars, inside clusters that form part 
of a larger region of coherent star formation in our Galaxy. The study 
of clusters in a wide age range, from 4 to 30 Myr is of interest for  
analysing the possible sequential star formation and the resulting age spread 
inside the complex. Discussion of the age properties and of the members' 
spatial distribution and NIR properties is postponed to our next paper.

Some different sources of uncertainty that, among others, might be affecting 
our membership results are: 

1. The $(U-B),(B-V)$ relation for PMS stars, may differ from the one for MS 
stars, and the reddening law for PMS stars, where unquantified amounts of 
remaining circumstellar dust and gas may be present, can also be different 
from the reddening relations describing extinction for MS stars (Delgado et al.
1998). 

2. The values of the reddening slopes may indeed change between clusters 
(Sagar \& Cannon 1997; Guetter \& Turner 1997).

3. The eventual presence of high reddening makes it harder to distinguish 
PMS members and non-member stars (Chini \& Wargau 1990).

4. The uncertainties in the theoretical isochrones, which arise both from the
model computation and from their translation to the observational plane as 
well.

It is not possible to evaluate the total uncertainty in the final 
classification due to these factors. We do, however, have arguments to 
support its reliability. 
First, we recall the restrictive conditions for the stars to be selected
as members, which require not only a simultaneously consistent location in all 
CM diagrams, but also a good photometric accuracy in all five $UBVRI$ bands. 

On the other hand, we have performed some follow-up spectroscopic studies of 
previously observed clusters, which provide results consistent with the 
photometric classification (Delgado et al. 1998, 1999, 2004). In the case of 
NGC\,2362 (Delgado et al. 2006), a satisfactory agreement is reached between 
the described membership assignment and expectations from independently 
observed properties of the stars, such as H$\alpha$ emission and 
X-ray activity. We think that the effects of all the above-mentioned sources
of uncertainty is decreased in the presence of not very high reddening, as is 
decidedly the case for NGC\,2362, and also -\,although not so favourable\,- as
for the clusters in the present sample. 

Finally, we recall the comparison of our membership assignments with those by 
Baume et al. (2003) for NGC\,3293, which shows a high degree of agreement: 
89\% of their 292 proposed members, also included in our observations, have 
some assignment by our procedure (either MS member or PMS member in at least 
one CMD). 

\section{Catalogue and conclusions}

Multicolour $UBVR_{C}I_C$ photometry has been obtained for 26962 stars in the 
fields of 10 young clusters, 9 of them located in the Carina star formation 
complex of the Milky Way. Ages of the clusters, as estimated from their MS and
post-MS probable members, range between LogAge(yr) 6.6 and 7.5. Matching of
the pixel coordinates with the 2MASS catalogue provides absolute coordinates 
for DAY-I catalogue and $JHK$ photometry for the stars in common.

Analysis of the joint evidence from the CC and CM diagrams, together with 
the comparison of the observational CM diagrams to theoretical PMS and post-MS
isochrones, led to estimates of both MS and PMS membership for individual 
stars in every cluster field. A total of 842 stars have been selected as PMS 
members 
in the different clusters, with respect to at least one of the three sets of
isochrone models used to estimate PMS membership. For some stars, membership 
assignment and corresponding age values were obtained according to these three 
models.

For the first time, a catalogue of probable PMS cluster members has been 
elaborated. This catalogue is available at the CDS (Centre de Donn\'ees 
astronomiques de Strasbourg) at {\tt http://cdsweb.u-strasbg.fr/CDS-f.html}, 
and it contains the basic 
information from the calculations and analysis described here. The column 
contents is as follows. Column 1: a 15-character string with the star 
identification. The first 9 characters are the IAU cluster number, and the 
last five characters are the identification number in our photometric files. 
Columns 2 and 3 list the right ascension and declination in degrees for Epoch 
2000; Columns 4-8 the optical photometric values $V$, $(U-B)$, $(B-V)$, 
$(V-R)$, and 
$(V-I)$; Columns 9-11 the NIR photometric values $K$, $(J-H)$, and $(H-K)$; 
Column 12 the membership indication: 1, PMS member assignments in three CM 
diagrams; 2, PMS member assignments in all four CM diagrams. Column 13 
indicates absorption $A_V=3.1 \times E(B-V)$ when available. Columns 14-16
show PMS ages with respect to the S00, P99, and D97 models, respectively. In 
Table\,4 we list the first 10 lines of the catalogue as an example, which 
correspond to stars in NGC\,2367.

\begin{table*}
\caption{Catalogue of PMS candidates}
\label{table:1}
     $$ 
         \begin{array}{l c c c r r r r c r r c c c c c}
            \hline
            \noalign{\smallskip}
OCL-Star~No. & RA   & Dec  & V & U-B & B-V & V-R & V-I & K & J-H & H-K & Mem & A_V &     & LogAge (yr) &      \\
               & 2000 & 2000 &   &     &     &     &     &   &     &     &     &    & S00 &     P99     & D97  \\
            \noalign{\smallskip}
            \hline
            \noalign{\smallskip}
C0718$-$218$-$00366 & 109.950218 & -21.908415 & 19.09 &      & 1.77 & 1.06 & 2.07 & 14.48 & 0.61 &  0.07 & 1 &      & 6.26 & 6.18 &      \\
C0718$-$218$-$00409 & 109.956973 & -21.893997 & 18.58 &      & 1.25 & 0.74 & 1.49 & 16.11 & 0.33 & -0.69 & 1 &      & 7.00 & 7.00 &      \\
C0718$-$218$-$00529 & 109.975359 & -21.915802 & 18.83 &      & 1.52 & 0.92 & 1.85 & 14.67 & 0.67 &  0.05 & 1 &      & 6.72 & 6.63 & 6.00 \\
C0718$-$218$-$00608 & 109.988120 & -21.880320 & 18.67 &      & 1.48 & 0.92 & 1.79 & 14.63 & 0.54 &  0.33 & 1 &      & 6.68 & 6.60 & 6.00 \\
C0718$-$218$-$00618 & 109.991959 & -21.989092 & 14.76 & 0.51 & 0.56 & 0.34 & 0.72 & 13.12 & 0.19 & -0.08 & 1 &      & 6.85 &      &      \\
C0718$-$218$-$00628 & 109.992614 & -21.905472 & 18.17 &      & 1.57 & 0.96 & 1.86 & 13.90 & 0.70 &  0.19 & 1 &      & 6.11 & 6.00 &      \\
C0718$-$218$-$00635 & 109.993885 & -21.946789 & 15.16 & 0.72 & 1.32 & 0.80 & 1.52 & 11.57 & 0.48 &  0.09 & 2 & 1.76 & 6.08 &      &      \\
C0718$-$218$-$00647 & 109.995650 & -21.917452 & 17.75 &      & 1.52 & 0.94 & 1.83 & 13.61 & 0.62 &  0.11 & 1 &      & 6.08 & 6.00 &      \\
C0718$-$218$-$00656 & 109.998575 & -21.938709 & 15.75 & 0.84 & 1.49 & 0.90 & 1.74 & 11.65 & 0.57 &  0.02 & 2 & 2.32 & 6.08 &      &      \\
C0718$-$218$-$00761 & 110.013962 & -21.901218 & 15.10 & 0.56 & 0.75 & 0.47 & 0.95 & 12.95 & 0.06 &  0.03 & 2 & 2.11 & 6.60 &      &      \\
           \noalign{\smallskip}
            \hline
         \end{array}
     $$

\end{table*}

The catalogue is meant as a starting point for more detailed studies of PMS
stars, either individually or as members of hierarchically-structured stellar 
systems. In particular, we consider it as a basis for follow-up spectroscopic 
analysis and eventual observations in other wavelength ranges, mainly in the
NIR range.  

\begin{acknowledgements}
This work has been supported by the Spanish MCYT through grant
AYA2004-05395 and by the Consejer\'\i a~ de Educaci\'on y Ciencia de la Junta
de Andaluc\'\i a, through TIC\,101.
We made use of the NASA ADS Abstract Service and of the WEBDA data base, 
developed by Jean-Claude Mermilliod at the Laboratory of Astrophysics of the 
EPFL (Switzerland), and further developed and maintained by Ernst Paunzen 
at the Institute of Astronomy of the University of Vienna (Austria). 
This publication makes use of data products from the Two Micron All
    Sky Survey, which is a joint project of the University of Massachusetts
    and the Infrared Processing and Analysis Center/California Institute of
    Technology, funded by the National Aeronautics and Space Administration
    and the National Science Foundation.
\end{acknowledgements}


\begin{thebibliography}{}

\bibitem[Ahumada et al. 2001]{} Ahumada, A.V., Clari\'a, J.J., Bica, E., 
Dutra, C.M., \& Torres, M.C., 2001, \aap, 377, 845

\bibitem[Alfaro et al. 1991]{} Alfaro, E. J., Cabrera-Ca\~no, J., \& Delgado, 
A. J., 1991, \apj, 378, 106

\bibitem[Alfaro et al. 1992]{} Alfaro, E. J., Cabrera-Ca\~no, J., \& Delgado, 
A. J., 1992, \apj, 399, 576

\bibitem[Baade 1983]{} Baade, D., 1983, \aaps, 51, 235

\bibitem[Balog \& Kenyon 2002]{bal02} Balog, Z. \& Kenyon, S., 2002, \aj, 124, 
2083

\bibitem[Balona \& Laney]{bal96} Balona L.A., and Laney C.D. 1996, \mnras, 
281, 1341

\bibitem[Baume et al. 2003]{} Baume, G., V\'azquez, R., Carraro, G., \& 
Feinstein, A., 2003, \aap, 402, 549 [B03]

\bibitem[Baume, V\'azquez, \& Feinstein 1999]{bau99} Baume, G., V\'azquez, R. 
A., \& Feinstein, A., 1999, \aaps, 137, 233

\bibitem[Bailyn et al. 1999]{} Bailyn, C., DePoy, D., Agostinho, R., 
Mendez, R., Espinoza, J., and Gonzalez, D. 1999, The First Year of Operations 
of the YALO Consortium, AAS 195, 8706

\bibitem[Bessell \& Brett 1988]{} Bessell, M.S., \& Brett, J.M., 1988, PASP,
100, 1134

\bibitem[Brandner et al. 2001]{} Brandner, W., Grebel, E. K., Barb\'a, R. H., 
Walborn, N. R., \& Moneti, A., 2001, \aj, 122, 858

\bibitem[Carpenter 2001]{} Carpenter, J. M., 2001, \aj, 121, 2851

\bibitem[Carraro \& Patat 2001]{} Carraro, G., \& Patat, F., 2001, \aap, 379, 
136

\bibitem[Carraro, Patat, \& Baumgardt 2001]{car01} Carraro, G., Patat, F., \&
Baumgardt, H., 2001, \aap, 107, 114

\bibitem[Claria 1976]{cla76} Claria, J.J., 1976, \aj, 81, 155

\bibitem[D'Antona \& Mazzitelli]{} D'Antona, F., \& Mazzitelli, I., 1997, in  
``Cool stars in Clusters and Associations'', eds. R. Pallavicini \& G. Micela, 
MemSAIt, 68, n.4, 807.

\bibitem[DeGioia-Eastwood et al. 2001]{deg01} DeGioia-Eastwood, K., Throop,
H., Walker, G., \& Cudworth, K. M., 2001, \apj, 549, 578

\bibitem[Delgado et al. 1998]{del98} Delgado, A. J., Alfaro, E. J., Moitinho, 
A., \& Franco, J., 1998, \aj, 116, 1801

\bibitem[Delgado et al. 1999]{del99} Delgado, A. J., Miranda, L. F., \& 
Alfaro, E. J., 1999, \aj, 118, 1759

\bibitem[Delgado \& Alfaro 2000]{del00} Delgado, A. J. \& Alfaro, E. J., 2000,
\aj, 119, 1848 

\bibitem[Delgado et al. 2004]{del04} Delgado, A. J., Miranda, L. F., 
Fern\'andez, M., \& Alfaro, E. J., 2004, \aj, 128, 330

\bibitem[Delgado et al. 2006]{del04} Delgado, A. J., Gonz\'alez-Mart\'\i n, O.,
Alfaro, E. J., \& Yun, J., 2006, \apj, 646, 269

\bibitem[de Winter et al. 1997]{dew97} de Winter, D., Koulis, C., Th\'e, P. S.,
van den Ancker, M. E., P\'erez, M. R., \& Bibo, E. A., 1997, \aaps, 121, 223

\bibitem[Feinstein \& Marraco 1980]{fei80} Feinstein, A., \& Marraco, H.G., 1980, PASP, 92, 266

\bibitem[Feinstein, Marraco, \& Forte 1976]{fei76} Feinstein, A., Marraco, H.G., 
\& Forte, J.C., 1976, \aaps, 24, 389

\bibitem[Fernandes, Lebreton, \& Baglin 1996]{fer96} Fernandes, J., Lebreton, Y., \& 
Baglin, A., 1996, \aap, 311, 127

\bibitem[FitzGerald et al. 1978]{fit78} FitzGerald, M. P., Kackson, P. D., 
Luiken, M., Grayzeck, E. J., \& Moffat, A. F. J., 1978, \mnras, 182,
607

\bibitem[Flaccomio et al.(2006)]{2006A&A...455..903F} Flaccomio, E., 
Micela, G., \& Sciortino, S.\ 2006, \aap, 455, 903 

\bibitem[Forbes \& DuPuy 1978]{for78} Forbes, D. \& DuPuy, D., 1978, \aj, 83, 
266

\bibitem[Forte \& Orsatti 1984]{for84} Forte, J. C. \& Orsatti, A. M., 1984
\apjs, 56, 211

\bibitem[Girardi et al. 2002]{} Girardi, L., Bertelli, G., Bressan, A., 
Chiosi, C., Groenewegen, M.A.T., Marigo, P., Salasnich, B., and Weiss, A. 
 2002, \aap, 391, 195

\bibitem[de Grijs et al. 2002]{} de Grijs, R., Gilmore, G.F., Mackey, A.D., 
Wilkinson, M.I., Beaulieu, S.F., Johnson, R.A., \& Santiago, B.X., 2002, 
\mnras, 337, 597.

\bibitem[Guetter \& Turner 1997]{} Guetter, H.H., \& Turner, D.G., 1997, \aj,
113, 2116 

\bibitem[H\"agele et al. 2003]{} H\"agele, G.F., Barb\'a, R.H., \& Bosch, G.L., 2003, 
BAAA, 46, 52

\bibitem[Haisch, Lada, \& Lada 2001]{hai01} Haisch Jr., K. E., Lada, E. A., \&
Lada, C. J., 2001, \apj, 553, L153

\bibitem[]{} Hartmann,  L., 2005. In "Star Formation in the Era of Three Great 
Observatories". Cambridge, MA, 2005. 
{\tt http://cxc.harvard.edu/stars05/agenda/program.html}

\bibitem[Herbig 1998]{her98} Herbig, G.H., 1998, \apj, 497, 736

\bibitem[Herbst \& Miller 1982]{hmi82} Herbst, W., \& Miller, D.P., 1982, 
AJ, 87, 1478

\bibitem[Hern\'andez et al. 2005]{her05} Hern\'andez, J., Calvet, N., 
Hartmann, L., Brice\~no, C., Sicilia-Aguilar, A., \& Berlind, P., 2005, \aj,
129, 856

\bibitem[Heske \& Wendker 1984]{hes84} Heske, A. \& Wendker, H., J., 1984, 
\aaps, 57, 205

\bibitem[Hillenbrand et al. 1993]{hil93} Hillenbrand, L. A., Massey, P., Strom,
S. E., \& Merrill, K. M., 1993, \aj, 106, 1906

\bibitem[Hillenbrand \&  Hartmann 1998]{} Hillenbrand, L.A., \&  Hartmann, 
L.W., 1998, \apj, 492, 540 

\bibitem[Jeffries \& Oliveira 2005]{jef05} Jeffries, R. D. \& Oliveira, J. M,
2005, \mnras, 358,13

\bibitem[Kharchenko et al. 2005]{} Kharchenko, N.V., Piskunov, A.E., R\"oser, S., 
Schilbach, E., \& Scholz, R.-D. 2005, \aap, 438, 1163

\bibitem[Kenyon \& Hartmann 1995]{} Kenyon, S.J., \& Hartmann, L., 1995, \
apjs, 101,117

\bibitem[Lada(2005)]{2005ccsf.conf..141L} Lada, E.~A.\ 2005, Cores to 
Clusters: Star Formation with Next Generation Telescopes, 141 

\bibitem[Lada \& Alves(2004)]{2004IAUS..221....3L} Lada, E.~A., \& Alves, 
J.~F.\ 2004, IAU Symposium, 221, 3 

\bibitem[Lawson, Lyo, \& Muzzerolle 2004]{law04} Lawson, W. A., Lyo, A-R., \& 
Muzzerolle, J., 2004, \mnras, 351, L39

\bibitem[Luhman et al. 2003]{luh03} Luhman, K. L., Stauffer, J. R., Muench, A. 
A., Rieke, G. H., Lada, E. A., Bouvier, J., \& Lada, C. J., 2003, \apj, 593, 
1093

\bibitem[Lyo, Lawson, \& Bessell 2004]{lyo04} Lyo, A-R., Lawson, W. A., \& 
Bessell, M. S., 2004, \mnras, 355,363
 
\bibitem[Mamajek, Meyer, \& Liebert 2002]{mam02} Mamajek, E. E., Meyer, 
M. R., \& Liebert, J., 2002, \aj, 124, 1670

\bibitem[Marco, Bernabeu, \& Negueruela 2001]{mar01} Marco, A., Bernabeu, G., 
\& Negueruela, I., 2001, \aj, 121,2075

\bibitem[Marco \& Negueruela 2002]{mar01} Marco, A.\& Negueruela, I., 2002, 
\aap, 393,195

\bibitem[Marschall, Comins, \& Karshner 1990]{mar90} Marschall, L. A., Comins, 
N. F., \& Karshner, G. B., 1990, \aj, 99, 1536

\bibitem[Massey et al. 2001]{} Massey, P., DeGioia-Eastwood, K., \& 
Waterhouse, E., 2001, \aj, 121, 1050

\bibitem[Moffat \& Vogt 1973]{mof73} Moffat, A.F.J., \& Vogt, N., 1973, \aaps, 10, 135

\bibitem[Moffat \& Vogt 1975]{mof75} Moffat, A.F.J., \& Vogt, N., 1975, \aaps, 20, 125

\bibitem[Munari, Carraro, \& Barbon 1998]{mun98} Munari, U., Carraro, G., \&
Barbon, R., 1998, \mnras, 297, 867

\bibitem[Palla \& Staller 1999]{} Palla, F., \& Staller, S.W., 1999, \apj, 525, 772

\bibitem[Palla et al. 2005]{pal05} Palla, F., Randich, S., Flaccomio, E., and 
Pallavicini, R. 2005, \apj, 626, L49

\bibitem[Park et al. 2000]{par00} Park, B.-G., Sung, H., Bessell, M. S., \&
Kang, Y. H., 2000, \aj, 120, 894

\bibitem[Park \& Sung 2002]{par02} Park, B.-G. \& Sung, H., 2002, \aj, 123, 892

\bibitem[Patat \& Carraro 2001]{pat01} Patat, F. \& Carro, G., 2001, \mnras,
325, 1591

\bibitem[Pozzo et al. 2003]{poz03} Pozzo, M., Naylor, T., Jeffries, R. D., \&
Drew, J. E., 2003, \mnras, 341, 805

\bibitem[Prisinzano et al. 2005]{pri05} Prisinzano, L., Damiani, F., Micela, 
G., \& Sciortino, S., 2005, \aap, 430, 941

\bibitem[Raboud et al. 1997]{rab97} Raboud, D., Cramer, N., \& Bernasconi, 
P.A., 1997, \aap, 325, 167

\bibitem[Raboud \& Mermilliod 1998]{rab98} Raboud, D., \& Mermilliod, J.C., 
1998, \aap, 333, 897

\bibitem[Sagar \& Cannon 1997]{} Sager, R., \& Cannon, R.D., 1997, \aaps, 122, 9

\bibitem[Sagar \& Joshi 1979]{sag79} Sagar, R. \& Joshi, U. C., 1979, \apss, 
66, 3

\bibitem[Sagar \& Joshi 1981]{sag81} Sagar, R. \& Joshi, U. C., 1981, \apss, 
75, 465

\bibitem[Sanner et al. 2001]{} Sanner, J., Brunzendorf, J., Will, J.-M., \& Geffert, M., 
2001, \aap, 369, 511

\bibitem[Sartori, L\'epine,\& Dias 2003]{sar03} Sartori, M. J., L\'epine, J. 
R. D., \& Dias, W. S., 2003, \aap, 404, 913 

\bibitem[Sicilia-Aguilar et al. 2005]{sic05} Sicilia-Aguilar, A., Hartmann, L. 
W., Hern\'andez, J., Brice\~no, C., \& Calvet, N., 2005, \aj, 130, 188

\bibitem[Shobbrook 1983]{} Shobbrook, R.R., 1983, \mnras, 205, 1215

\bibitem[Siess, Dufour \& Forestini]{} Siess, L., Dufour, E., \& Forestini, 
M., 2000, \aap, 358, 593 

\bibitem[Slesnick, Hillenbrand, \& Massey 2002]{sle02} Slesnick, L. S., 
Hillenbrand, L. A, \& Massey, P., 2002, \apj, 576, 880

\bibitem[Stauffer at al. 1989]{stau89} Stauffer, J., Hartmann, W. L., Jones,
B. F., \& McNamara, B. R., 1989, \apj, 342, 285

\bibitem[Stone 1988]{sto88} Stone, R. C., 1988, \aj, 96, 1389

\bibitem[Sung, Besell, \& Lee 1997]{sun97} Sung, H., Bessell, M. S., \& Lee, 
S.-W., 1997, \aj, 114, 2644

\bibitem[Sung, Besell, \& Lee 1998]{sun98} Sung, H., Bessell, M. S., \& Lee, 
S.-W., 1998, \aj, 115, 734

\bibitem[Sung, Chun, \& Besell 2000]{sun00} Sung, H., Chun, M.-Y., \& Bessell, 
M. S., 2000, \aj, 120, 333

\bibitem[Th\'e et al. 1985]{the85} Th\'e, P. S., Hageman, T., Westerlund, B.E.,
\& Tjin A Djie, H. R. E., 1985, \aap, 151, 391

\bibitem[Trullols \& Jordi 1997]{tru97} Trullols, E. \& Jordi, C., 1997, \aap, 
324, 549

\bibitem[Turner 1994]{} Turner, D.G., 1994, RMAA, 29, 163

\bibitem[Turner et al. 1980]{} Turner, D.G., Grieve, G.R., Herbst, W., \& Harris, W.E.,
1980, \aj, 85, 1193

\bibitem[Turner \& Moffat 1980]{} Turner, D.G., \& Moffat, A.F.J., 1980, \mnras, 192, 283

\bibitem[van den Ancker et al. 1997]{van97} van den Ancker, M. E., Th\'e, P. 
S., Feinstein, A., V\'azquez, R. A., de Winter, D., \& P\'erez, M. R., 1997, 
\aaps, 123, 63

\bibitem[van den Ancker, Th\'e, \& de Winter 2000]{van00} van den Ancker, M. 
E., Th\'e, P. S., \& de Winter, D., 2000, \aap, 362, 580

\bibitem[Vazquez et al. 1994]{} V\'azquez, R.A., Baume, G., Feinstein, A., \& Prado, P.,
1994, \aaps, 106, 339

\bibitem[Vazquez \& Feinstein 1990]{} V\'azquez, R.A.,\& Feinstein, A., 1990, \aaps, 86,209

\bibitem[Vazquez \& Feinstein 1991]{} V\'azquez, R.A.,\& Feinstein, A., 1991, \aaps, 87,383

\bibitem[Verschueren, Hensberge, \& De Loore 1990]{ver90} Veschueren, W., 
Hensberge, H., \& De Loore, C., 1990, \apss, 170, 245

\bibitem[Vogt \& Moffat 1972]{} Vogt, N., \& Moffat, A.F.J., 1972, \aaps, 7, 133 

\bibitem[Walker 1961]{wal61} Walker, M. F., 1961, \apj, 133, 43

\bibitem[Wesselink 1969]{} Wesselink, A.J., 1969, \mnras, 146,329

\bibitem[Witt, Schild, \& Kraiman 1984]{wit84} Witt, A. N., Schild, R., E., \&
Kraiman, J. B., 1984, \apj, 281, 708

\end{thebibliography}
\end{document}